\documentclass[11pt]{article}

\usepackage{amsmath}
\usepackage{amsfonts}
\usepackage{amssymb}
\usepackage{graphicx}
\usepackage{setspace}
\usepackage{authblk}
\usepackage{cite}
\usepackage{setspace}
\usepackage{array,multirow}
\usepackage{subcaption}
\usepackage[skip=0.5ex]{subcaption}

\usepackage{float}
\usepackage[dvipsnames]{xcolor}
\usepackage{hyperref}
\hypersetup{colorlinks,linkcolor={BrickRed},citecolor={RoyalBlue},urlcolor={RoyalBlue},filecolor={Turquoise}}

\newcommand*\rot{\rotatebox{90}}

\begin{document}
	
	\title{\textbf{Heavy Higgs boson Searches at the LHC in the light of a Left-Right Symmetric Model}}	
	
	\author[]{Sanchari Bhattacharyya\footnote{sanchari1192@gmail.com}}
	\affil[]{\emph{University of Calcutta} \\ \emph{\emph{92} Acharya Prafulla Chandra Road, Kolkata 700009}}
	\date{}
	
	\maketitle
	
	\vskip 2cm

	\begin{abstract}
		We investigate a Left-Right symmetric model respecting $SU(3)_C \otimes SU(2)_L \otimes U(1)_L \otimes SU(2)_R \otimes U(1)_R$ local gauge symmetry. We study the interactions of the heavy neutral and charged scalars of this model along with their production at the hadron collider and their subsequent decays. We analyze the collider searches of two heavy scalars, one of them is charge neutral and another one is singly charged. In both the cases we consider their associated production at the Large Hadron Collider (LHC) and finally concentrate only on the leptonic final states. We perform both cut-based and multivariate analysis using Boosted Decision Tree algorithm for 14 TeV as well as 27 TeV LHC run with 3000 fb$^{-1}$ integrated luminosity. As expected, the multivariate analysis shows a better signal-background discrimination compared to the cut-based analysis. In this article, we show that a charged Higgs of mass 750 GeV and 1.2 TeV can be probed with $2.77 \sigma$ ($4.58 \sigma$) and $1.38 \sigma$ ($3.66 \sigma$) significance at 14 (27) TeV run of LHC.
	\end{abstract}

	\section{Introduction}
	
	It is well known that Standard Model (SM) of particle physics has been extremely successful in describing the interactions of the elementary particles. The discovery of Higgs boson at Large Hadron Collider (LHC), CERN \cite{higgs_atlas, higgs_cms} has added another feather in its cap. Despite of being so successful, it is still unable to explain some of the natural phenomena which are already experimentally established, for example the explanation of Dark Matter (DM) or tiny neutrino mass etc. It is also unknown to us that whether the discovered Higgs boson is the only scalar candidate in nature or there are also other scalars with heavier masses which are similarly responsible for Electroweak Symmetry Breaking (EWSB). All of these unexplained facts actually motivate the physicists to look beyond SM (BSM).
	
	In the existing literature, there are several studies which actually deal with the phenomenology of extended Higgs sector \cite{Higgs-review}. Many of them have argued that the idea of one Higgs boson is not complete and there may be other representations also which may give rise to other required Higgs bosons having a heavier or lighter mass compared to the SM Higgs boson. We are hopeful that with the advancement of technologies a detailed study about the properties of SM Higgs boson, for example its decays, branching ratios (BR), its couplings, precision measurements \cite{higgs-precision1,higgs-precision2} will be possible which will make the picture of the scalar sector clear. The Higgs self-coupling could not be precisely measured yet. To constrain this value, people also have searched for di-Higgs at the collider \cite{higgs-precision2,dihiggs}. Extended Higgs sector may also have some bearings on this dark matter sector, Higgs mass hierarchy or neutrino mass issues. In some models, singlet scalar has been considered as a suitable DM candidate \cite{singlet-DM}. The presence of charged Higgs may contribute to the radiative masses of neutrino \cite{numass}. In Left-Right symmetric models (LRSM), people have studied about the mass generation of neutrinos with help of extended triplet or singlet Higgs bosons \cite{triplet-neutrinomass,Nu_mass1}. Additional Higgs bosons can also play a crucial role in dealing with the flavor problems \cite{2hdm}. 
	The non-detection of such a scalar from the direct searches from the LHC pushes the exclusion limits on the masses of such scalars to higher and higher scales.
	
	In quest of such a complete theory, we investigate a model which respects $SU(3)_C \otimes SU(2)_L \otimes U(1)_L \otimes SU(2)_R \otimes U(1)_R$ ($32121$) \cite{32121} local gauge symmetry. This can be obtained via a two-step symmetry breaking from $E_6$ \cite{E6} Grand Unified group with $\left[ SU(3)_C \otimes SU(3)_L \otimes SU(3)_R \right]$ as the intermediate step. We shall only be interested in the Left-Right (LR) symmetric gauge group, 32121 and in the phenomenology of its scalar sector. This model contains the fermions from the full \textbf{27}-plet of $E_6$, among them 11 are heavy BSM fermions. Two of these heavy fermions, one being Dirac like and another being Majorana like, are suitable DM candidates. The detailed analysis regarding Dark Matter aspects of this model has been discussed in \cite{32121DM}. 
	%{\color{red} This model gives rise to a two component DM scenario. One of the DM candidates has a larger rate of interaction compared to the other. The relic particle with the larger interaction rate, satisfies the constraints from Direct detection experiments when a dimension-6 effective four-fermion interaction is introduced with a new coupling strength. The other DM candidate, with a smaller interaction rate is able to satisfy relic density constraints only when the coannihilation channels between these two relic candidates are opened up. Thus together they present a promising DM scenario and one can constrain the parameter spaces using the recent results of the direct detection of Dark Matter and relic density measurements from PLANCK collaboration.}
	Apart from the SM gauge bosons the gauge sector comprises of three heavy BSM gauge bosons. The scalars in 32121 arise from the ($\textbf{1}, \textbf{3}, \bar{\textbf{3}}$) representation of $SU(3)_C \otimes SU(3)_L \otimes SU(3)_R$. They are color singlet and heavy scalars. One of them must have similar properties like SM Higgs boson. 
	At this point it is worth making a note that though this 32121 model has a few similarities with extensively studied Left-Right Symmetric Model (LRSM) \cite{lrsm} but it offers some unique features providing interesting phenomenology which makes this model significant and different from LRSM. 
	\begin{itemize}
		\item Unlike LRSM which mainly deals with the extension of scalar and gauge sector, the 32121 model contains a few charged and neutral heavy BSM fermions from \textbf{27}-plet of $E_6$ along with extended Higgs and gauge sector. Some of these fermions offer interesting phenomenology which we have discussed in ref. \cite{32121}.
		\item The particle sector of model already contains two Dark Matter candidates which satisfy the limits from direct detection experiments and relic density measuments \cite{32121DM}.
		\item Unlike LRSM, 32121 model contains a heavy stable charged particle which may leave charged tracks in the detector. So pair production of such field leaves two charged tracks producing an unique signature of this model.
		\item 32121 contains heavy stable neutral scalar fields also. So production of such a field along with a heavy stable charged particle will leave a single charged track with missing transverse energy. This also offers an unique signature which is not so discussed in other models.
		\item Along with an usual heavy gauge boson present in several models including LRSM $Z'$, a new heavy neutral gauge boson $A'$ is present in this model. It couples to the SM fermions and have similar properties like $Z'$ but has a different mass which we also could derive using experimenal results from LHC.
	\end{itemize} 
	Some of the BSM Higgs bosons show interesting signatures at High Luminosity-LHC (HL-LHC). In this article we shall mainly analyse the properties of some of the heavy Higgs bosons and their signatures at the LHC with 14 and 27 TeV high luminosity run. 
	
	In this article we plan to describe the model breifly in section \ref{sec2} where we mainly discuss the particle sector of this model with special emphasis on the scalar sector of our interest. We also discuss the properties and production mechanisms of the some BSM scalars including heavy neutral and singly charged scalars at the LHC. In section \ref{sec3} we perform the signal-background phenomenology of these two BSM Higgs bosons considering the cut-based analysis as well as multivariate analysis. We shall see that the signal-background discrimination is much better in case of multivariate analysis. Finally we conclude in section \ref{sec4}.

	\section{Description of 32121 Model}
	\label{sec2}
	
	We start with the Left-Right (LR) symmetric 32121 gauge group. A two-step symmetry breaking of $E_6$ can lead to 32121, though we will not be interested in this specific symmetry breaking pattern. This model is rich in particles which are listed in Table \ref{table1} with their corresponding gauge quantum numbers. In this article, among all the particles we will mainly study the interactions of some of the scalars which may generate interesting signatures at hadron collider.
	
	The Higgs multiplets present in the Table \ref{table1} are instrumental in breaking down $SU(3)_C \otimes SU(2)_L \otimes U(1)_L \otimes SU(2)_R \otimes U(1)_R$ to the $SU(3)_C \otimes SU(2)_L \otimes U(1)_Y$ and then to $SU(3)_C \otimes U(1)_{EM}$. $L$ and $R$ denote Left and Right repectively. One can calculate the electric charge, $Q$ as, $Q = T_{3L} + T_{3R} + Y_L/2 + Y_R/2$ where $Y_L/2$ and $Y_R/2$ are the $U(1)_{L,R}$ hypercharges and are noted down in the last two columns of Table \ref{table1} respectively. $T_{3L}$ and $T_{3R}$ are the third component of weak isospin corresponding to $SU(2)_L$ and $SU(2)_R$ gauge group respectively.
	
	\begin{table}[h!]
		\centering
		\begin{tabular}{|c|c|lrccc|}
			\hline \hline
			& Fields & & Gauge & quantum & numbers & \\ [1.0ex]
			\hline \hline
			& & $3_C$ & $2_L$ & $2_R$ & $1_L$ & $1_R$ \\ [1.0ex]
			\hline 
			& $L_L$ & $1$ & $2$ & $1$ & $-1/6$ & $-1/3$ \\ [1.0ex]
			& $\bar{L}_R$ & $1$ & $1$ & $2$ & $1/3$ & $1/6$ \\ [1.0ex]
			& $\bar{L}_B$ & $1$ & $2$ & $2$ & $-1/6$ & $1/6$ \\ [1.0ex]
			Fermions & $\bar{l}_S$ & $1$ & $1$ & $1$ & $1/3$ & $-1/3$ \\ [1.0ex]
			& $Q_L$ & $3$ & $2$ & $1$ & $1/6$ & $0$ \\ [1.0ex]
			& $\bar{Q}_R$ & $\bar{3}$ & $1$ & $2$ & $0$ & $-1/6$ \\ [1.0ex]
			& $\bar{Q}_{LS}$ & $\bar{3}$ & $1$ & $1$ & $-1/3$ & $0$ \\ [1.0ex]
			& $Q_{RS}$ & $3$ & $1$ & $1$ & $0$ & $1/3$ \\ [1.5ex]
			\hline				
			& $\Phi_B$ & $1$ & $2$ & $2$ & $1/6$ & $-1/6$ \\ [1.0ex]
			Scalar & $\Phi_L$ & $1$ & $2$ & $1$ & $1/6$ & $1/3$  \\ [1.0ex]
			Fields & $\Phi_R$ & $1$ & $1$ & $2$ & $-1/3$ & $-1/6$ \\ [1.0ex]
			& $\Phi_S$ & $1$ & $1$ & $1$ & $-1/3$ & $1/3$ \\ [1.5ex]
			\hline				
			& $G^i ,\; i=1,...,8$& $8$ & $1$ & $1$ & $0$ & $0$ \\ [1.0ex]
			& $W^i_{L}, i=1,2,3$ & $1$ & $3$ & $1$ & $0$ & $0$ \\ [1.0ex]
			Gauge fields & $W^i_{R}, i=1,2,3$ & $1$ & $1$ & $3$ & $0$ & $0$ \\ [1.0ex]
			& $B_L$ & $1$ & $1$ & $1$ & $0$ & $0$ \\ [1.0ex]
			& $B_R$ & $1$ & $1$ & $1$ & $0$ & $0$ \\ 
			\hline
		\end{tabular}
		\caption{Fermions and Bosons in $32121$ model with their respective gauge quantum numbers}
		\label{table1}
	\end{table}

	\subsection{Gauge sector}
	
	The gauge sector of 32121 model has two charged and four neutral gauge bosons. In the charged sector, one has been identified with the SM $W$ boson and the other field is the heavy $W'$ boson. In the neutral gauge sector two fields have been identified with SM $Z$ and photon. The remaining two fields (after symmetry breaking) are denoted as $Z'$ and $A'$. The masses and mixings along with the interactions in electro-weak gauge sector are controlled by the four gauge coupling constants, $g_{2L}$, $g_{2R}$, $g_{1L}$ and $g_{1R}$ along with the vacuum expectation values (vevs) of the scalar fields. If one follows the symmetry breaking pattern of $SU(2)_R \otimes U(1)_L \otimes U(1)_R$ to $U(1)_Y$, one can have an expression like,
	\begin{equation}
	\label{GCValue1}
	\dfrac{1}{g_Y^2} = \dfrac{1}{g_{2R}^2} + \dfrac{1}{g_{1L}^2} + \dfrac{1}{g_{1R}^2}
	\end{equation} where $g_Y$ denotes the $U(1)_Y$ gauge coupling constant. $g_{2L}$ is identified with the $SU(2)_L$ gauge coupling constant of SM, $g$. We have chosen $g_{2L}=g_{2R} = g$ and $g_{1L}=g_{1R}$ to keep our Lagrangian Left-Right symmetric. With these choices one can fix the gauge parameters of the 32121 model. On the other hand, the lower limits of the vevs of the Higgs fields can be fixed from the experimental lower limits of the heavy gauge bosons. A deltailed study on the gauge sector of 32121 model can be found in \cite{32121}.\\
	
	\subsection{Fermion sector}
	
	As already mentioned, in 32121 model we have 27 fermions. Their chiral components are as follows,
	\begin{eqnarray}
	L_{L} &=& \begin{pmatrix}
	\nu_{L} \\ e_{L}
	\end{pmatrix}, \hspace{1.2cm} \hspace{1.3cm} L_{R} = \begin{pmatrix}
	\nu_{R} \\ e_{R}
	\end{pmatrix} \nonumber \\
	Q_{L} &=& \begin{pmatrix}
	u_{L} \\ d_{L}
	\end{pmatrix}, \hspace{2.4cm} Q_{R} = \begin{pmatrix}
	u_{R} \\ d_{R}
	\end{pmatrix} \nonumber \\
	Q_{LS} &=& q_{SL}, \hspace{0.5cm} Q_{RS} = q_{SR}, \hspace{0.5cm} l_S \hspace{0.5cm} \mbox{and,}\nonumber \\
	L_B &=& \begin{pmatrix}
	N_1 & E_1 \\ E_2 & N_2
	\end{pmatrix}  \hspace{0.3cm} \mbox{and} \hspace{0.7cm}
	\tilde{L}_B = \begin{pmatrix}
	N_2^c & E_2^c \\ E_1^c & N_1^c
	\end{pmatrix}
	\end{eqnarray}
	$L_{L,R}$ and $Q_{L,R}$ contain the SM leptons and quarks respectively along with a right-handed neutrino. Rest of fields are BSM fermions. $Q_{LS}$ and $Q_{RS}$ form a four-component Dirac-like color triplet quark whereas $N_1,~N_2^c$ and $E_1,~E_2^c$ construct neutral and singly charged Dirac-like lepton $N$ and $E$ respectively. $l_S$ is color singlet and carries equal and opposite $U(1)_{L,R}$ hypercharges. $l_S$ and $l_S^c$ together form a Majorana-like neutral fermion $L_S$.
	
	The interactions between the Higgs fields and the fermions are responsible for the masses of the fermions. The relevant Yukawa Lagrangian is as follows.
	\begin{eqnarray}
	\label{LRYukawa}
	\mathcal{L}_{Y} &=& y_{qij} \bar{Q}_{iL} \Phi_B Q_{jR} + \tilde{y}_{qij} \bar{Q}_{iR} \tilde{\Phi}_B Q_{jL} + y_{lij} \bar{L}_{iL} \Phi_B L_{jR} + \tilde{y}_{lij} \bar{L}_{iR} \tilde{\Phi}_B L_{jL} \nonumber \\ 
	&+& y_{sij} \bar{Q}_{iLS} \Phi_S Q_{jRS} + y_{LBij} \; Tr \left[ \bar{L}_{iB} \tilde{L}_{jB} \right] \Phi_S^c + \frac{y_{LSij}}{\Lambda} \bar{l}_{iS} l_{jS}^c \Phi_S \Phi_S \nonumber \\
	&+& y_{BBij}\; Tr\left[ \bar L_{iB} \tilde \Phi_B \right] l_{jS}^c  + y_{ijBR} \bar{L}_{iL} L_{jB} \Phi_R + y_{ijBL} \bar{L}_{iR} L_{jB}^\dagger \tilde{\Phi}_L \nonumber \\
	&+& y_{ijLRS} \bar{Q}_{iL} Q_{jRS}^* \tilde{\Phi}_L + y_{ijRLS} \bar{Q}_{iR} Q_{jLS}^* \tilde{\Phi}_R + h.c.
	\end{eqnarray}
	\normalsize
	where, $i,j = 1,2,3$ are generation numbers and $y$(s) are Yukawa coupling constants. $\Phi_S^*$ is complex conjugate of $\Phi_S$, $\tilde{\Phi}_B = \sigma_2 \Phi_B^* \sigma_2$ and $\tilde{L}_B=\sigma_{2} L_B^* \sigma_{2}$. 
	
	The first line of Eq. \ref{LRYukawa} shows the terms generating the masses of the SM fermions. The terms present in the second line of Eq. \ref{LRYukawa} are responsible for giving masses to the heavy BSM fermions. The Yukawa coupling matrices are non-diagonal in general. But for the sake of simplicity, we have not gone through any detailed calculations regarding mixings among the SM and BSM fermions. It is to be noted that we have written a dimension-5 term to generate the Majorana mass of $l_S$ where $\Lambda$ is an arbitrary mass scale which is assumed to arise from any higher energy scale than the LR symmetry breaking scale. The rest of the terms represent the mixings among BSM and SM fermions. Here we note that, we can write only Dirac-like mass term for the neutrino in our model. The $U(1)_{L,R}$ hypercharges of $L_L$ and $L_R$ restrict ourselves from writing any Majorana-mass term for neutrino. 
	
	\subsection{Scalar sector of 32121}
	
	There are several scalar fields in this model. The Higgs fields which are mainly responsible for the symmetry breaking from $32121 \longrightarrow SU(3)_C \otimes SU(2)_L \otimes U(1)_Y \longrightarrow SU(3)_C \otimes U(1)_{EM}$ are, one Higgs bi-doublet ($\Phi_B$), one left-handed ($\Phi_L$), one right-handed ($\Phi_R$) weak doublets and a singlet Higgs boson ($\Phi_S$). $\Phi_S$ is $SU(2)$ singlet but carries {equal and opposite} $U(1)$ hypercharges. {It is to be noted that being a pure singlet under SM gauge group, $\Phi_S$ does not take part in electroweak symmetry breaking.} These color singlet scalars arise from (\textbf{1}, \textbf{3}, $\bar{\textbf{3}}$) representation of the Trinification gauge group ($\left[ SU(3)_C \otimes SU(3)_L \otimes SU(3)_R \right]$). Among these fields, $\Phi_R$ is instrumental in breaking the LR symmetry. The allignment of the Higgs felds are as following.
	
	\begin{eqnarray}
	\Phi_B &=& \begin{pmatrix}
	\frac{1}{\sqrt{2}}(k_1 + h_1^0 + i \xi_1^0)  & h_1^+  \\ h_2^- & \frac{1}{\sqrt{2}}(k_2 + h_2^0 + i \xi_2^0)
	\end{pmatrix},  \nonumber \\
	\vspace{1cm}
	\Phi_L &=& \begin{pmatrix}
	h_L^+ \\ \frac{1}{\sqrt{2}}(v_L + h_L^0 + i \xi_L^0)
	\end{pmatrix}, %\hspace{3cm}
	\Phi_R = \begin{pmatrix}
	\frac{1}{\sqrt{2}}(v_R + h_R^0 + i \xi_R^0) \\ h_R^- 
	\end{pmatrix}, \Phi_S = \frac{1}{\sqrt{2}}(v_S + h_S^0 + i \xi_S^0)
	\end{eqnarray}
	
	The Higgs potential of the 32121 model, $\cal V$ is composed of two parts, ${\cal V}_1$ and ${\cal V}_2$. It is given by,
	\begin{eqnarray}
	\label{pot}
	\mathcal{V}_1 = &-&\mu_1^2 Tr \left( {\Phi_B}^{\dagger} \Phi_B\right) - \mu_3^2 \left( {\Phi_L}^{\dagger} \Phi_L + {\Phi_R}^{\dagger} \Phi_R \right)  - \mu_4^2 {\Phi_S}^{\dagger} \Phi_S   \nonumber \\
	&+& \lambda_1 Tr \left[ ({\Phi_B}^{\dagger} \Phi_B)^2\right] + \lambda_3 \left( Tr\left[ {\Phi_B}^{\dagger} \tilde{\Phi}_B\right]  Tr\left[ \tilde{\Phi}_B^{\dagger} \Phi_B\right] \right) \nonumber \\
	&+& \alpha_1 (\Phi_S^{\dagger} \Phi_S)^2 + \beta_1 Tr\left[ {\Phi_B}^{\dagger} \Phi_B\right]  (\Phi_S^{\dagger} \Phi_S) + \gamma_1 \left[ (\Phi_L^{\dagger} \Phi_L) + (\Phi_R^{\dagger} \Phi_R)\right] (\Phi_S^{\dagger} \Phi_S) \nonumber \\
	&+& \rho_1 \left[ (\Phi_L^{\dagger} \Phi_L)^2 + (\Phi_R^{\dagger} \Phi_R)^2\right] 
	%+ \rho_2 \left[ (\Phi_L^{\dagger} \Phi_R \Phi_R^{\dagger} \Phi_L) + (\Phi_R^{\dagger} \Phi_L \Phi_L^{\dagger} \Phi_R) \right]  
	+ \rho_3 \left[ (\Phi_L^{\dagger} \Phi_L) (\Phi_R^{\dagger} \Phi_R)\right] + c_1 Tr\left[ {\Phi_B}^{\dagger} \Phi_B\right] \left[ (\Phi_L^{\dagger} \Phi_L) + (\Phi_R^{\dagger} \Phi_R)\right]   \nonumber \\
	&+& c_3 \left[  ( \Phi_L^{\dagger} \Phi_B  \Phi_B^{\dagger} \Phi_L ) + ( \Phi_R^{\dagger} \Phi_B^{\dagger} \Phi_B  \Phi_R ) \right] + c_4 \left[ ( \Phi_L^{\dagger} \tilde{\Phi}_B \tilde{\Phi}_B^{\dagger} \Phi_L ) + ( \Phi_R^{\dagger} \tilde{\Phi}_B^{\dagger} \tilde{\Phi}_B \Phi_R ) \right]
	\end{eqnarray}
	and, \begin{equation}
	\label{tri-linear}
	\mathcal{V}_2 = \mu_{BS} Tr \left[ {\Phi^\dagger _B} \tilde \Phi_B\right]  \Phi_S^\ast + h.c.
	\end{equation}
	
	The parameters in $\mathcal{V}$ are considered to be real. $\mathcal{V}$ is also LR symmetric and obeys the gauge symmetry of 32121 model. %In the above,  $\tilde{\Phi}_B \equiv \sigma_2 \Phi_B^* \sigma_{2}$.
	
	Apart from the above symmetries, $\mathcal{V}_1$ is also symmetric under the global phase transformations like, 
	\begin{equation}
	\Phi_B \rightarrow e^{i \theta_B}\; \Phi_B; ~~\Phi_L \rightarrow e^{i \theta_L}\; \Phi_L; ~~\Phi_R \rightarrow e^{i \theta_R}\; \Phi_R {~\rm and} 
	~\Phi_S \rightarrow e^{i \theta_S}\; \Phi_S.
	\label{global}
	\end{equation}
	Whereas, the terms present in $\mathcal{V}_2$ explicitly breaks this symmetry. Now, if we choose both $k_1$ and $k_2$ to be non-zero, the terms proportional to $\lambda_3$ in $\mathcal{V}_1$ give rise to some bilinear terms like $h_1^0 h_2^0$, $h_1^+ h_2^-$ which makes $\mathcal{V}_1$ break the aforementioned global symmetry spontaneously.
	This results into emergence of a massless mode in the physical spectra of particles. This massless scalar, also called Majoron \cite{heeck} couples to the $Z$ boson thus contributes to the invisible decay width of $Z$. Invisible decay width of $Z$, being very accurately measured quantity, would basically disfavour such a massless scalar.
	This issue of getting unwanted Goldstone mode can be avoided in two ways. One simple option is to choose one of $k_1$ and $k_2$ to be zero which will make such bilinear terms (like $h_1^0 h_2^0$, $h_1^+ h_2^-$) vanish and turn the potential $\mathcal{V}_1$ invariant under such a global symmetry. Another way is to consider $\mathcal{V}_2$ in addition to $\mathcal{V}_1$ as the scalar potential. As $\mathcal{V}_2$ breaks the global symmetry explicitly, we can get rid of the extra massless mode in this way. In \cite{32121}, it is discussed in detail that the presence of $\mathcal{V}_2$ does not affect the masses and the mixings in the scalar sector in a significant way. Hence, we choose $k_2$ to be \emph{zero}. 
	%One could also consider this massless mode as a Majoron \cite{heeck}. In that case, in our model, this mode has a coupling with $Z$ bosons which will provide a tight constraint on its mass and couplings. In this article, this specific study is beyond our scope. So we choose to go with a different approach.

	A non-zero value of $v_R$ is necessary to lead the Left-Right symmetry breaking. Whereas, $v_S$ also needs to be non-zero as it is responsible for $U(1)$ symmetry breaking. A non-zero value of $v_L$ will along with a non-zero $v_R$ will again spontaneously break the global symmetry mentioned in Eq. \ref{global} which will give rise to an extra unwanted Goldstone mode. In order to avoid such a problem, we choose $v_L=0$ \cite{32121}.
	
	There are 10 real parameters in the scalar potential of this model, $\lambda_1$, $\lambda_3$, $\rho_1$, $\rho_3$, $c_1$, $c_3$, $c_4$, $\alpha_1$, $\beta_1$ and $\gamma_1$. We accept only those values of the quartic parameters which make the scalar potential bounded from below and which are allowed by the SM-Higgs signal strengths \cite{32121}. 
	
	Among all the scalar fields, there are five neutral CP-even scalar fields, $h^0$, $h_2^0$, $h_L^0$, $H_R^0$ and $H_S^0$. $h^0$ has been identified with the SM-Higgs. The neutral CP-odd scalar sector contains two physical fields, $\xi_2^0$ and $\xi_L^0$. In addition to these scalars, there are two charged Higgs fields, $H_1^\pm$ and $H_L^\pm$. $h_2^0$ and $\xi_2^0$ are mass degenerate at the tree level. In a similar fashion, $h_L^0$ and $\xi_L^0$ also have same mass. In this article, we will mainly concentrate on the scalars who belong to the Higgs bi-doublet, $\Phi_B$ and discuss their properties.\\\\
	
	$\bullet$ \textbf{\underline{Scalars from Bi-doublet Higgs field:}}
	\vspace{5mm}
	
	Apart from the SM-like Higgs, the bi-doublet Higgs field $\Phi_B$ comprises of some BSM scalar fields including two neutral CP-even ($h_2^0$) and CP-odd ($\xi_2^0$) scalars and a singly charged Higgs $H_1^\pm$. At tree level, the above scalar ($h_2^0$) and pseudoscalar ($\xi^0 _2$) have equal masses. With $k_2=0$,
	\begin{equation}
	m_{h_2^0}^2 = m_{\xi_2^0}^2 = \frac{1}{2}[ 4 \lambda_3 k_1^2  + (c_4 - c_3) v_R^2]
	\end{equation}
	The couplings of $h_2^0 ~(\xi_2^0)$ with a pair of scalars or a pair of gauge bosons are proportional to $k_2$.
	The \emph{zero} value of $k_2$ restricts $h_2^0 ~(\xi_2^0)$ to couple with a pair of other scalars or gauge bosons but they can have interactions with a pair of SM fermions (see Eq. \ref{LRYukawa}). The couplings of $h_2^0 ~(\xi_2^0)$ with a pair of SM fermions depend on $k_2$ as well as $k_1$. So even after choosing the \emph{zero} value of $k_2$ a non-vanishing coupling of $h_2^0 ~(\xi_2^0)$ with the SM fermions is possible. From Eq. \ref{LRYukawa} it is evident that the coupling of $h_2^0 ~(\xi_2^0)$ with the up quark sector is proportional to the bottom quark sector Yukawa coupling and vice-versa. This implies that the coupling of $h_2^0 ~(\xi_2^0)$ with a pair of bottom quarks is proportional to top Yukawa coupling (see Appendix A, Table \ref{coupling}). To find the limit on the mass of $h_2^0 ~(\xi_2^0)$ we have produced these heavy scalars in association with a pair of $b$-quarks with a further decay to $b$ quark pair. ATLAS and CMS have already performed a search for heavy neutral scalar which is produced in association with a pair of $b$ quarks at $\sqrt{s} = 13$ TeV \cite{Atlas_h2, Cms_h2}. Using this result, we compare $\sigma \times BR$ obtained in 32121 model with the measured rate by ATLAS Collaboration and find a lower limit on $m_{h_2^0} ~(m_{\xi_2^0})$. We find, $m_{h_2^0} ~(m_{\xi_2^0})$ must be greater than 800 GeV \cite{32121}.
	
\begin{figure}[H]
		\begin{center}
			\includegraphics[width=8.5cm, height=7cm]{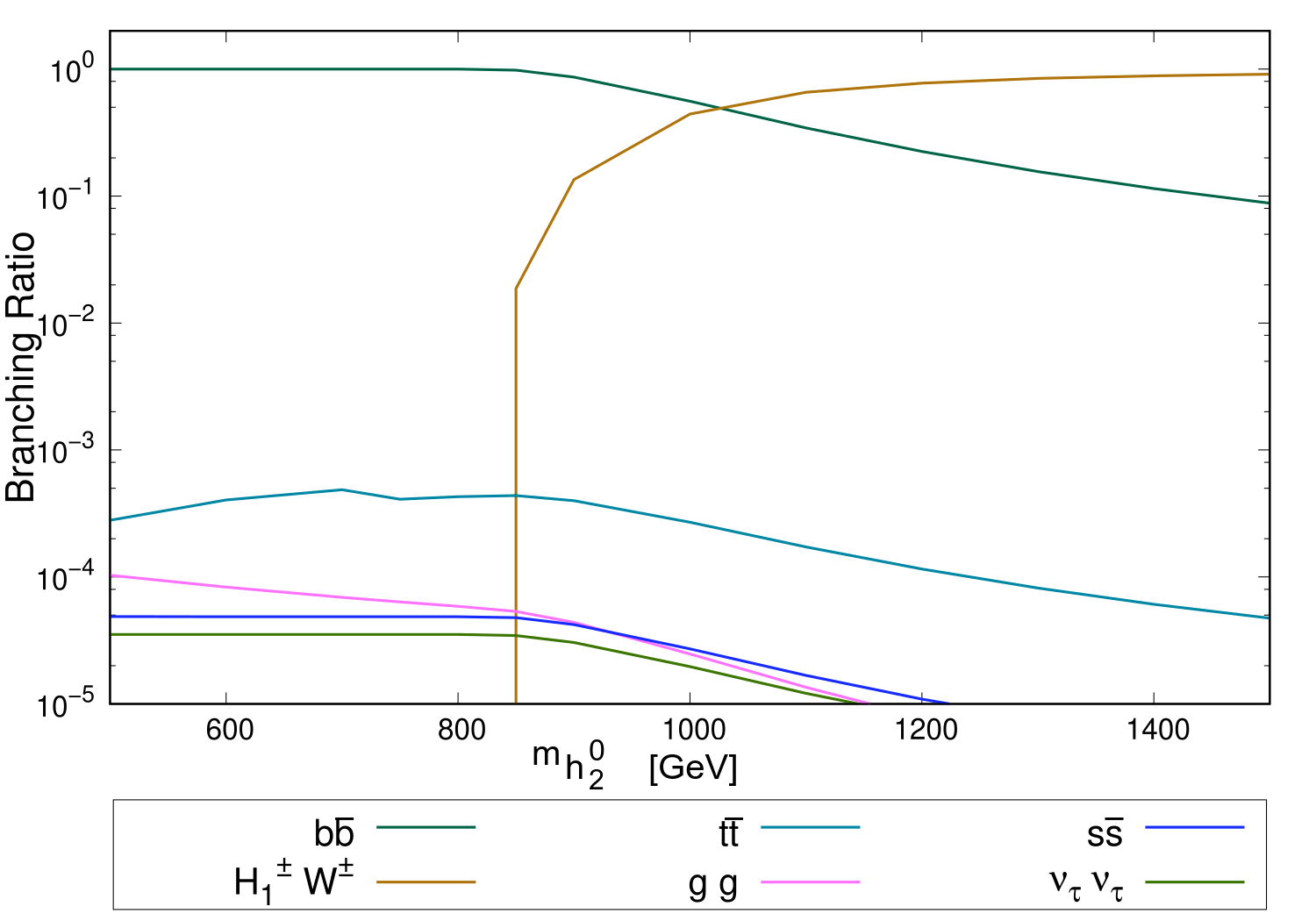}
			\includegraphics[width=8.5cm, height=7cm]{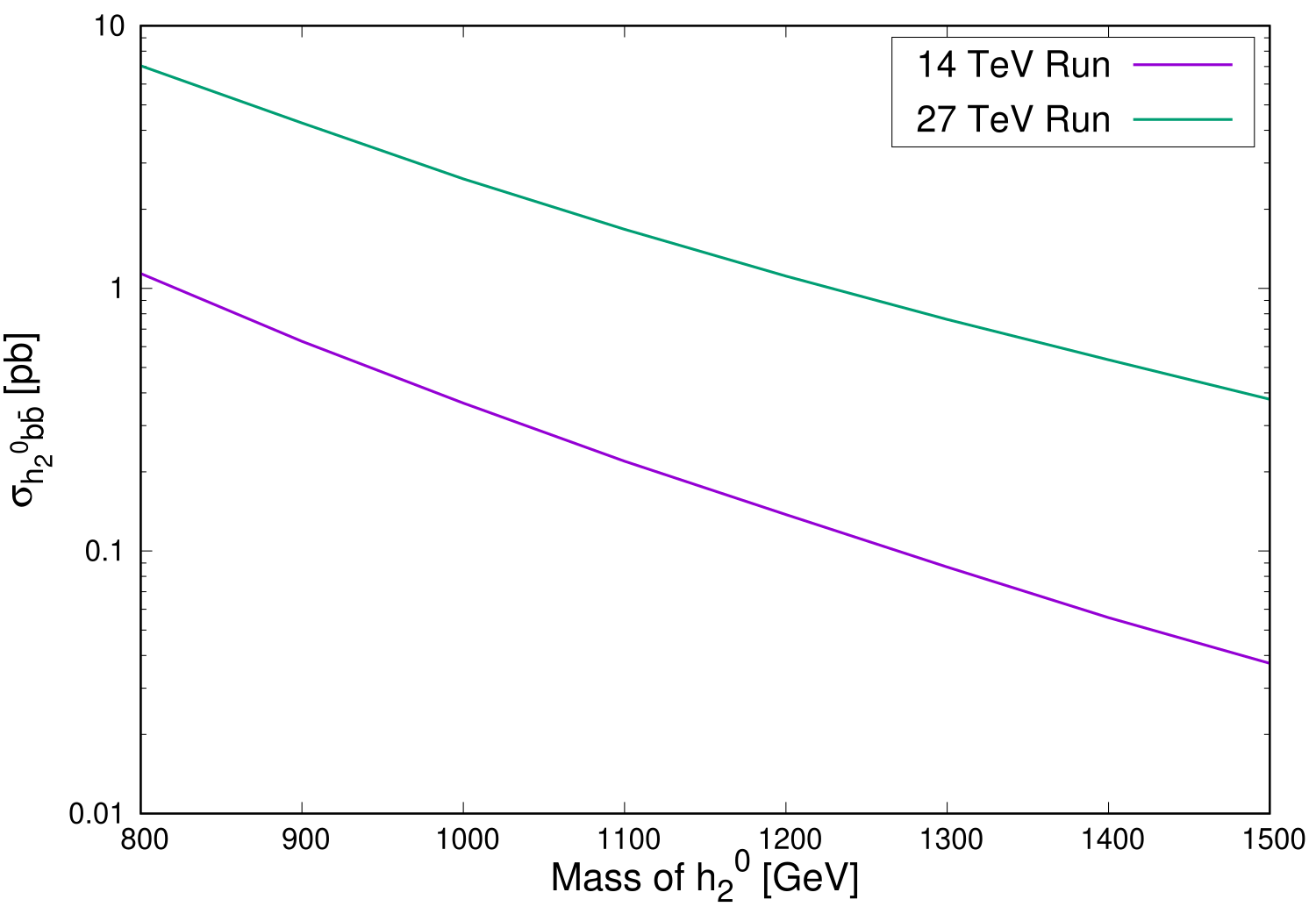}
			\caption{The left figure represents the branching ratios of $h^0_2 ~(\xi_2^0)$ to different final states. Here we set the mass of $H_1^\pm$ to 750 GeV. The figure on the right shows the associated production cross section of $h_2^0 ~(\xi_2^0)$  ($\sigma$) at the LHC at 14 and 27 TeV proton proton center of mass energy.}
			\label{h20-prod}
		\end{center}
	\end{figure}

	At the LHC, one of the dominating ways of producing $h_2^0 ~(\xi_2^0)$ is via gluon gluon fusion. Unlike SM Higgs, here a triangle loop of bottom quark will mainly control the production cross-section \cite{32121}. Another dominant way to produce $h_2^0 ~(\xi_2^0)$ at the hadronc collider is the associated Higgs production as previously dicussed. One can produce $h_2^0 ~(\xi_2^0)$ in association with two bottom quarks (see Fig. \ref{scalar_prod}). This large production cross-section will sensitively depend on the top Yukawa coupling. This in turn makes us consider the associated production mechanism while generating the heavy scalars at the collider.
	We present the associated production cross-section and decay branching ratios of $h_2^0 ~(\xi_2^0)$ in Fig. \ref{h20-prod}. It is to be noted that this production cross-section is at the partonic level without considering any showering or detector simulation.
	We note that $h_2^0 ~(\xi_2^0)$ has a dominant decay mode to $b\bar{b}$ untill the decay to $H_1^\pm W^\mp$ is kinematically allowed. In this plot the mass of $H_1^\pm$ has been set to 750 GeV.
	
	In order to generate such events, we have first implemented our model in \texttt{FeynRules2.0} \cite{feynrules} and then generated such processes using \texttt{Madgraph} \cite{madgraph} using \texttt{NNPDF3.0} parton distribution functions \cite{parton-dist}. We have also taken care of the QCD K-factor ($\sim 1.1$) following the ref. \cite{h2-qcd-k-factor, b_running}.
	
	Now, coming to the singly charged Higgs boson, $H_1^{\pm}$, it is another scalar field which is of our interest. $H_1^{\pm}$ has a mass,
	\begin{equation}
	m_{H_1^{\pm}}^2 = \frac{1}{2} (c_4-c_3) (k_1^2 + v_R^2)
	\end{equation}
	It can couple to SM fermions via Yukawa coupling (see Eq. \ref{LRYukawa}) and also has interactions with SM $W$ boson and heavy neutral scalar $h_2^0 ~(\xi_2^0)$. One dominant process of producing this charged scalar at the LHC is the production in association with a top and a bottom quark. Other mechanisms may include Drell-Yan process or even vector boson fusion process. ATLAS and CMS collaborations both have searched for heavy charged Higgs boson at 13 TeV run followed by a decay to a top and a bottom quark \cite{AtlasCharged1, AtlasCharged2, CmsCharged}. In our analysis, we have also produced $H_1^\pm$ in association with a top and a bottom with a further decay of $H_1^\pm$ again to a top and a bottom. We compare the event rates obtained in 32121 model with the result provided by ATLAS collaboration which $H_1^\pm$. We find, $m_{H_1^\pm} >$ 720 GeV \cite{32121}.
	
\begin{figure}[H]
		\begin{center}
			\includegraphics[width=8.5cm, height=8cm]{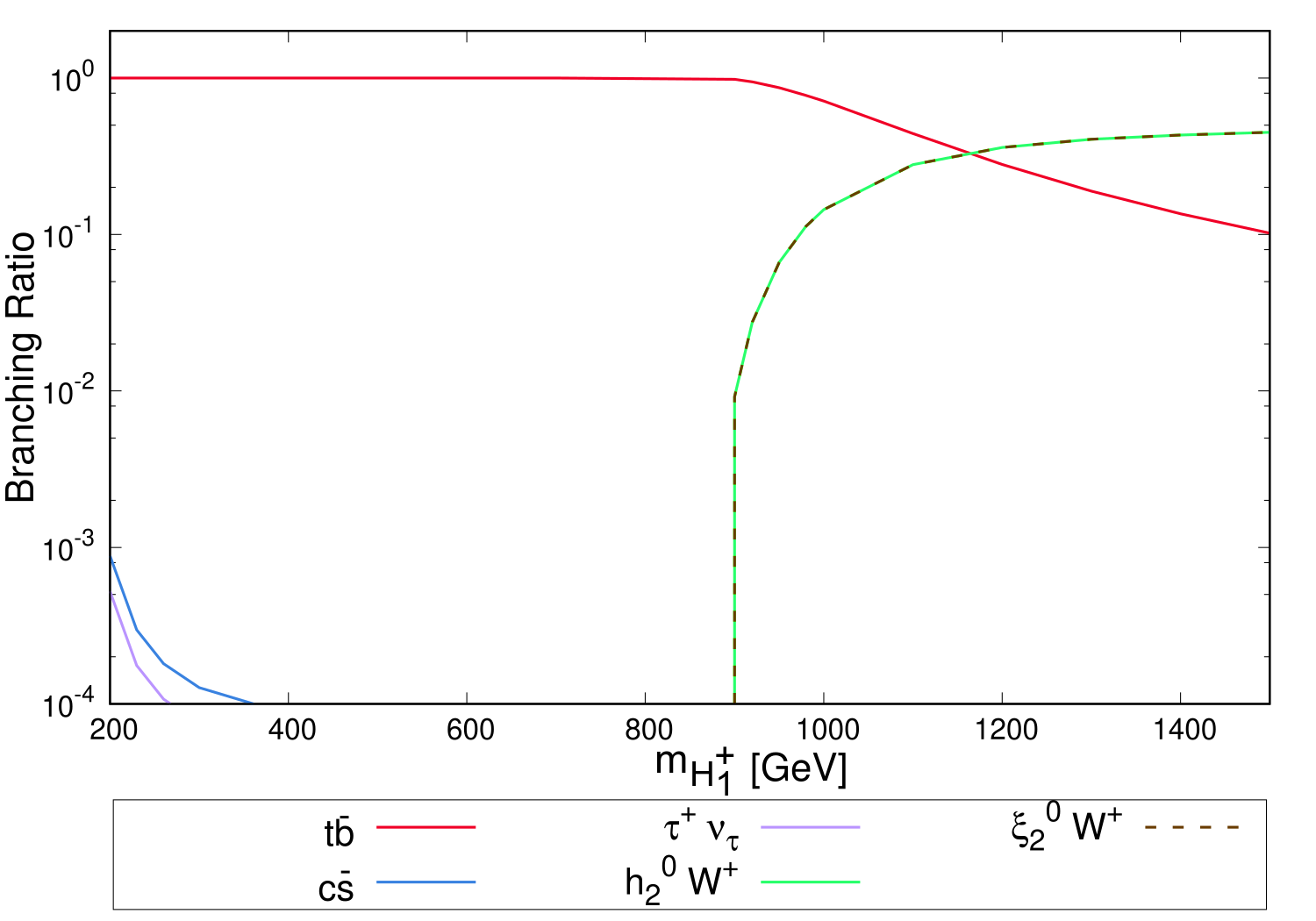}
			\includegraphics[width=8.5cm, height=8cm]{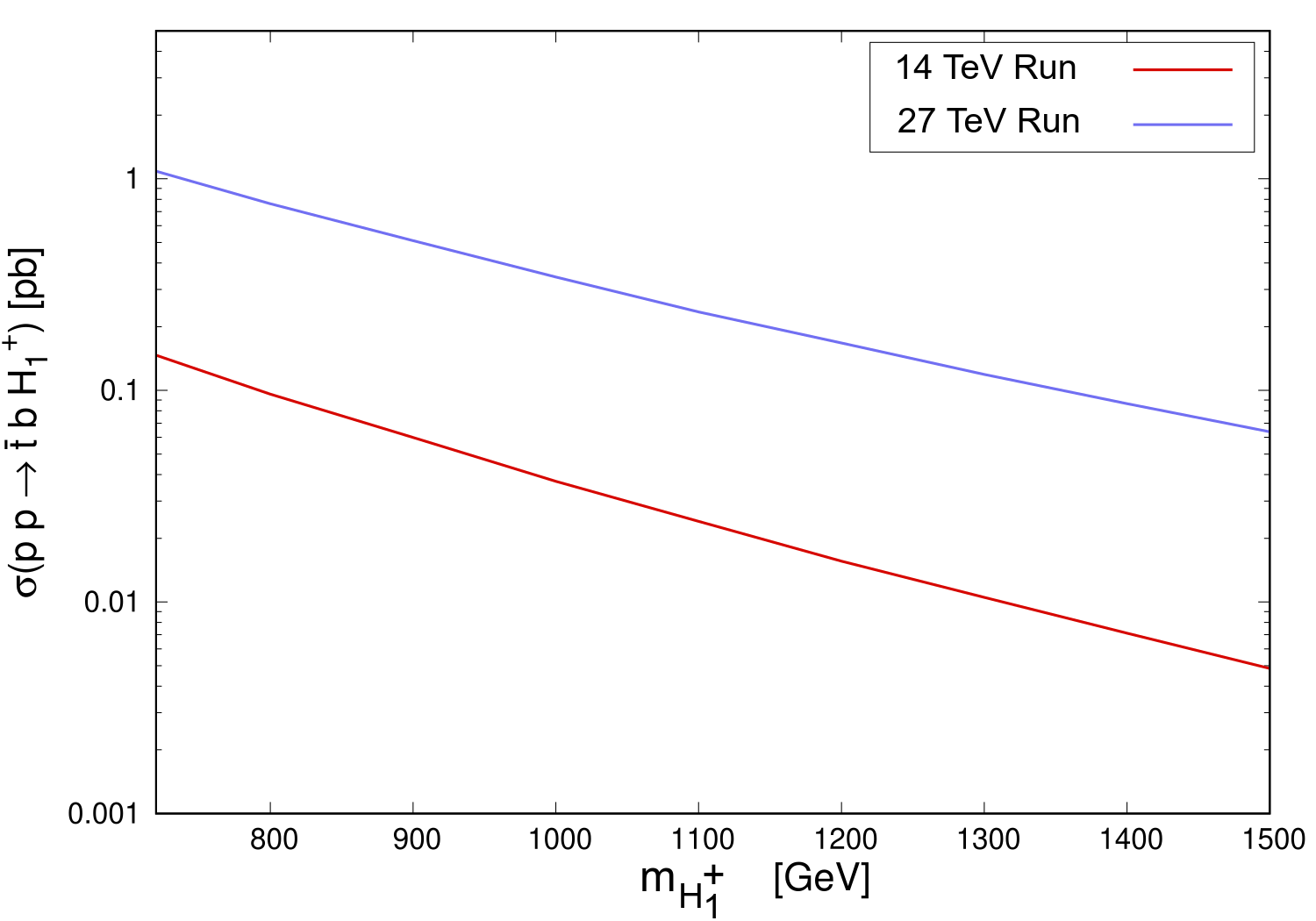}
			\caption{The figure on the left shows the branching ratios of $H_1^\pm$ to different final states where the mass of $h_2^0~(\xi_2^0)$ at its lowest limit ($800$ GeV). The right figure production cross-section ($\sigma$) of $H_1^\pm$ via $p p \rightarrow \bar{t} b H_1^+$ process at 14 and 27 TeV LHC run.}
			\label{h1p-prod}
		\end{center}
	\end{figure}

	While performing our analysis, we have considered to produce $m_{H_1^\pm}$ at the collider in association with $t~b$. The leading contribution will be from $g g \rightarrow \bar{t} b H_1$. In Fig. \ref{h1p-prod}, we present the production cross-section of $H_1^\pm$ at centre of mass energies of 14 TeV and 27 TeV along with the branching ratios of $H_1^\pm$ to different final states. We observe that $H_1^{\pm}$ mainly decays to a top and a bottom quark until it is allowed to decay to $h_2^0 ~(\xi_2^0) W^\pm$ kinematically. The  $H_1^+ \bar{t}b$ production cross-section varies from 0.15 (1) pb for $m_{H_1^\pm} = 720$ GeV to 0.005 (0.06) pb for $m_{H_1^\pm} = 1500$ GeV at 14 (27) TeV LHC run.
	
	In the next section, we will now present the signal-background study of $h_2^0 ~(\xi_2^0)$ and $H_1^\pm$ production at the LHC at 14 and 27 TeV run with 3000 fb$^{-1}$ integrated luminosity.
	
	\section{Signal-Background Analysis}
	\label{sec3}
	
	In the previous section we have already discussed about the production mechanisms and subsequent decays of the two scalars, $h_2^0 (~\xi_2^0)$ and $H_1^\pm$. The heavy neutral and charged scalars have some interesting decay channels to BSM fields. In this section we concentrate on the signal-background analysis of these two scalars at the LHC where we have considered such BSM decay channels of the heavy scalars. 
	
\begin{figure}[H]
		\begin{subfigure}{0.49\textwidth}
			\begin{center}
				\includegraphics[width=7cm, height=4.5cm]{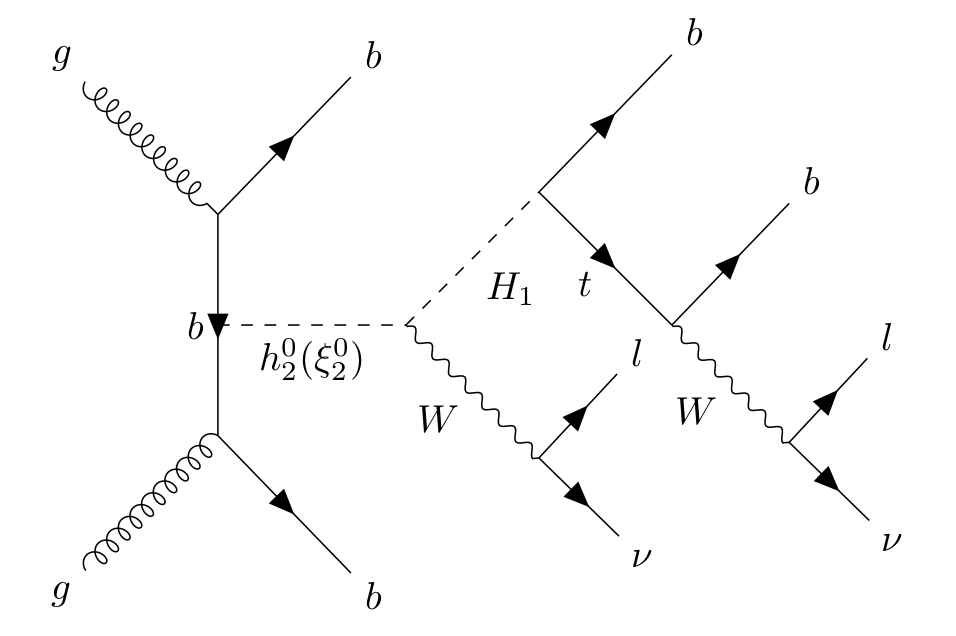}
			\end{center}
			\caption{}
		\end{subfigure}
		\begin{subfigure}{0.49\textwidth}
			\begin{center}
				\includegraphics[width=7cm, height=4.5cm]{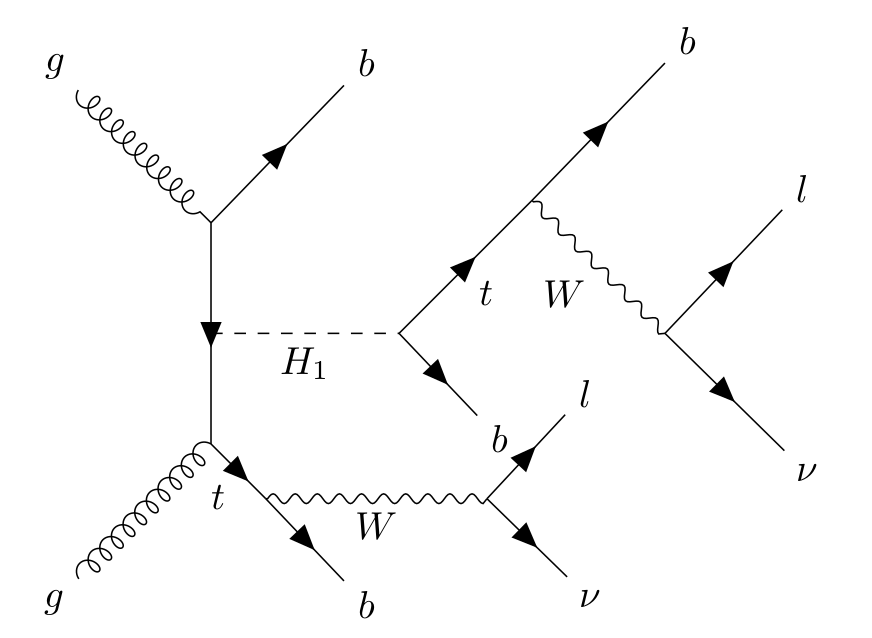}
			\end{center}
			\caption{}
		\end{subfigure}
		\caption{Feynman diagrams for the most dominant production processses of $h_2^0 ~(\xi_2^0)$ (a) and $H_1^\pm$ (b) at the LHC.}
		\label{scalar_prod}	
	\end{figure}

	One of the interesting channels to probe $h_2^0 (\xi_2^0)$ at the LHC is the following (see Fig. \ref{scalar_prod}).
	\begin{equation}
	p p \rightarrow h_2^0 (\xi_2^0) b \bar{b} \rightarrow (H_1^\pm W^\mp) b \bar{b} \rightarrow (t \bar{b} l^- \bar{\nu}_l) b \bar{b} \rightarrow b \bar{b} b \bar{b} l^+ l^- \nu_l \bar{\nu}_l  \nonumber
	\end{equation}
	
	Similarly to look for $H_1^\pm$ at the hadron collider, one may consider (see Fig. \ref{scalar_prod}),
	\begin{eqnarray}
	p p \rightarrow H_1^\pm t b \rightarrow (t \bar{b}) \bar{t} b \rightarrow (W^- \bar{b} b) W^+ b \bar{b} \rightarrow b \bar{b} b \bar{b} l^+ l^- \nu_l \bar{\nu}_l  \nonumber
	\end{eqnarray}
	We shall now briefly discuss these two channels with leptonic decay of $W$ boson with not-too-large background in the context of HL-LHC at 14 TeV and 27 TeV center of mass energy.
	
	Fig. \ref{scalar_prod} shows the leading order Feynman diagrams which are the most dominating for the production of heavy neutral $h_2^0 ~(\xi_2^0)$ and singly charged scalars $H_1^\pm$ at the LHC.
	We denote the production of $h_2^0 (\xi_2^0)$ and $H_1^\pm$ as signal 1 ($S_1$) and signal 2 ($S_2$) respectively. 
	Both the signals discussed above, have similar final states with four b jets, two oppositely charged leptons and missing transverse energy in the final state. This specific combination in the final state makes these signals unique as the chances of getting similar states coming from the Standard Model is quite low. 
	
	Among all the background processes, $t \bar{t}$ + jets production will be the most dominant. Other significant background effects include $b \bar{b} t \bar{t}$ production, $h t \bar{t}$ production, $Zt\bar{t}$ production, multijet processes etc.
	
	Initially we have set the transverse momentum (the component of momentum of a particle in the transverse plane, plane perpendicular to the beam axis) of b-tagged jet, light jets and leptons as, ${p_T}_b >$ 40 GeV, ${p_T}_j >$ 30 GeV, ${p_T}_l >$ 10 GeV respectively. We have also put an initial cut on missing transverse energy, $E_T \!\!\!\!\!\!/~~$ which is greater than 20 GeV.
	
	We perform this analysis for four chosen benchmark points corresponding to four different sets of masses of the scalars and their decay properties.
	 
	$S_1$ depends on the branching ratio of $h_2^0$ to $H_1^\pm W^\mp$ channel which is non-zero just after a certain mass of $h_2^0$ (see Fig. \ref{h20-prod}). Whereas $S_2$ depends on the $H_1^+ \rightarrow t \bar{b}$ branching ratio which is non-zero throughout the mass range of $H_1^\pm$ (see Fig. \ref{h1p-prod}) but this branching ratio reduces after the $H_1^\pm \longrightarrow h_2^0 ~(\xi_2^0) W^\pm$ decay channel opens up. In Table \ref{table2}, the four choices of the benchmark points have been presented.
	
	\begin{table}[h!]
		\centering
		\begin{tabular}{|c|c|c|c|}
			\hline \hline
			& $m_{h_2^0 (\xi_2^0)}$ [TeV] & $m_{H_1^\pm}$ [TeV] & Remarks \\[1.0ex]
			\hline \hline
			BP1 & 1 & 0.75 & Signal 1 $\longrightarrow$ On \\[1.0ex]
			& & & Signal 2 $\longrightarrow$ On \\[1.0ex]
			\hline
			BP2 & 0.8 & 0.75 & Signal 1 $\longrightarrow$ Off \\[1.0ex]
			& & & Signal 2 $\longrightarrow$ On \\[1.0ex]
			\hline
			BP3 & 1 & 1.2 & Signal 1 $\longrightarrow$ Off \\[1.0ex]
			& & & Signal 2 $\longrightarrow$ On \\[1.0ex]
			\hline
			BP4 & 1.2 & 1 & Signal 1 $\longrightarrow$ On \\[1.0ex]
			& & & Signal 2 $\longrightarrow$ On \\[1.0ex]
			\hline \hline
		\end{tabular}
		\caption{The four benchmark points considered in the analysis}
		\label{table2}
	\end{table} 
	
	For BP1 and BP4, the masses of $h_2^0 ~(\xi_2^0)$ and $H_1^\pm$ are such that both the signals, $S_1$ and $S_2$ are \emph{on} as the BR ($h_2^0 \rightarrow H_1^\pm W^\mp$) and BR ($H_1^+ \rightarrow t \bar{b}$) are non-zero, whereas for BP2 and BP3 the BR($h_2^0 \rightarrow H_1^\pm W^\mp$) is zero turning \emph{Signal 1} \emph{off}. 
	Furthermore, in case of BP1, for $S_1$ and $S_2$, $h_2^0 ~(\xi_2^0)$ and $H_1^+$ dominantly decay to $H_1^\pm W^\mp$ and $t\bar{b}$ respectively with the highest BR in the corresponding channels.
	For BP2, BR($H_1^+ \rightarrow t \bar{b}$) remains same as it is in case of BP1 keeping the \emph{Signal 2} unchanged. 
	However, in case of BP3 the $H_1^\pm$ dominantly decays to $h_2^0 W^\pm$ turning  BR($H_1^+ \rightarrow t \bar{b}$) small. 
	In a similar fashion for BP4 we get both signals \emph{on} as BP1 but with reduced branching ratios to corresponding channels and with reduced cross-sections.
	It is important to note here that the BP2 and BP3 practically imply the production of charged Higgs ($H_1^\pm$) only whereas BP1 and BP4 actually denote the production of both the scalars. 
	In Table. \ref{xsection} the production cross-sections corresponding to $S_1$ and $S_2$ as well as the branching ratio of $H_1^\pm$ and $h_2^0$ have been noted for all benchamark points.
	\begin{table}[H]
		\centering
		\begin{tabular}{|c|cc|cc|c|c|}
			\hline \hline
			& $\sigma_{S_1}$ & [fb] & $\sigma_{S_2}$ & [fb] & $BR(h_2^0 \rightarrow H_1^\pm W^\mp)$ & $BR(H_1^\pm \rightarrow tb)$ \\[1.0ex]
			\cline{2-5}
			& 14 TeV & 27 TeV & 14 TeV & 27 TeV &  &  \\[1.0ex]
			\hline \hline
			BP1 & 0.5521 & 4.451 & 0.7743 & 3.811 & 0.442 & 0.9998 \\[1.0ex]
			\hline
			BP2 & 0 & 0 & 0.7743 & 3.811 & 0 & 0.9998 \\[1.0ex]
			\hline
			BP3 & 0 & 0 & 0.2088 & 2.015 & 0 & 0.979 \\[1.0ex]
			\hline
			BP4 & 0.1916 & 1.828 & 0.3725 & 2.209 & 0.261 & 0.999 \\[1.0ex]
			\hline \hline
		\end{tabular}
		\caption{The production cross-section times the branching ratios in corresponding channels at 14 and 27 TeV LHC run are noted for $S_1$ ($\sigma_{S_1}$) and $S_2$ ($\sigma_{S_2}$). The values of the corresponding branching ratios important for $S_1$ and $S_2$ have also been noted.}
		\label{xsection}
	\end{table}

	\subsection{Cut-based Analysis}
	
\begin{figure}[H]
		\begin{center}
			\includegraphics[width=8.6cm, height=7cm]{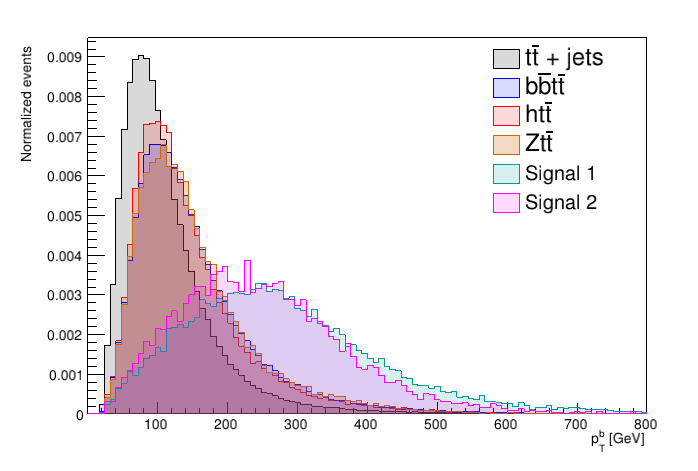}
			\includegraphics[width=8.6cm, height=7cm]{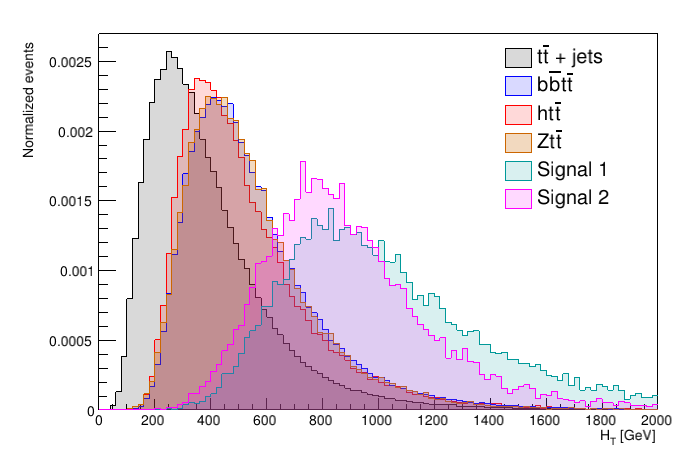}
			\caption{Distribution plots of transverse momentum, $p_T$ of leading b-tagged jet and scalar sum of $p_T$ of all the visible jets, $H_T$ of the signals and backgrounds for integrated luminosity 3000 fb$^{-1}$ at 14 TeV run of LHC.}
			\label{distplot_14}
		\end{center}
	\end{figure}

\begin{figure}[H]
		\begin{center}
			\includegraphics[width=8.6cm, height=7cm]{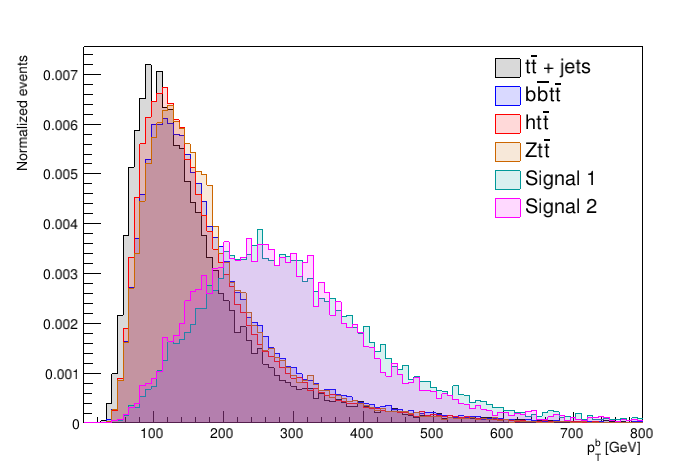}
			\includegraphics[width=8.6cm, height=7cm]{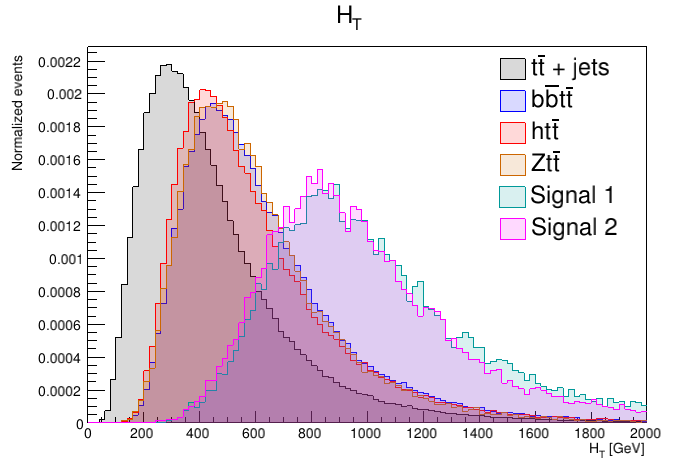}
			\caption{Distribution plots of transverse momentum, $p_T$ of leading b-tagged jet and scalar sum of $p_T$ of all the visible jets, $H_T$ of the signals and backgrounds for integrated luminosity 3000 fb$^{-1}$ at 27 TeV run of LHC.}
			\label{distplot_27}
		\end{center}
	\end{figure}

	In this section we present the signal-background analysis of $h_2^0 (\xi_2^0)$ and $H_1^\pm$ production at the LHC using cut-based approach. 
	After generating the signal and background events, we have passed them through \texttt{Pythia8} \cite{pythia8} already built in Madgraph to consider the showering and hadronization and used \texttt{Delphes3.5} \cite{delphes} for detector simulation. We demand there are \emph{at least} three b-tagged jets and \emph{at least} one charged lepton ($e$ or $\mu$) in the final state. Such a choice effectively turns down the number of events of multijet production which makes us ignore this background. 
	With these demands we have made, we plot the distributions of two important variables including the transverse momentum of leading b-tagged jet, $p_T^{b1}$ and the scalar sum of $p_T$ of all the visible jets, $H_T$ at 14 and 27 TeV HL-LHC run with 3000 fb$^{-1}$ integrated luminosity. In Figs. \ref{distplot_14} and \ref{distplot_27}, we present such distributions for the BP1 only. 
	In Figs. \ref{distplot_14} and \ref{distplot_27}, the distributions of $p_T$ of leading b-tagged jet and $H_T$ have been presented for each signal and background processes for BP1 at 14 and 27 TeV center of mass energy respectively with integrated luminosity 3000 fb$^{-1}$. The processes corresponding to the different color codes have been mentioned inside the plots. It is clearly understandable from the plots that an appropriate cut on $p_T$ of leading b-tagged jet and $H_T$ in each case can effectively reduce the background events compared to the signal events. Here we want to note that, for the other benchmark points the disributions of the signals are not significantly different compared to what is shown for BP1. We have optimized the cuts in such a way so that the significance of the signal does not vary significantly for all of the benchmark points. \\\\

	$\bullet$ \textbf{Event Selection}
	
	\vspace{3mm}
	
	As already mentioned, we choose to keep such events where there are at least three b-tagged jets in the final state. Using the information we get from the distribution plots in Figs. \ref{distplot_14} and \ref{distplot_27}, we apply and try to optimize the cuts on the variables we have considered i.e., $p_T^b$ and $H_T$ so that we can reduce certain amount of background events keeping the signal events as much as possible. In other words we try apply our cuts on the suitable variable in such a way so that we obtain maximum significance, $\mathcal{S}$ where $\mathcal{S}$ is given by,
	\begin{equation}
	\mathcal{S} = \sqrt{2\left[(S+B) ~ln \left(\dfrac{S+B}{B}\right)-S\right]}
	\label{significance}
	\end{equation}
	S and B stand for the number of signal and background events respectively.
	
	In Table \ref{table3}, we show the optimized cut flows providing the maximum significance, (see Eq. \ref{significance}) for all of the benchmark points at 14 TeV HL-LHC run. Here we select only those events where the transverse momentum of leading b-jet, $p_t^b >$ 240 GeV and $H_T >$ 990 GeV. 
		
	\begin{table}[H]
		\centering
		%\small
		\begin{tabular}{|c|c|c|c|c|c|c|c|}
			\hline \hline	
			& & & $N_b \ge 3$ & $p_{Tb1} > 240$ GeV & $H_T ~~> 990$ GeV & $S=N_{S1}+N_{S2}$ & $\mathcal{S}$ \\ [1.0ex]
			\hline \hline				
			& & $t\bar{t} + jets$ & 5127095 & 624321 & 209137 & $-$ & $-$ \\ [1.0ex]
			Backgrounds & & $b\bar{b}t\bar{t}$ & 40239 & 6228 & 1856 & $-$ & $-$ \\ [1.0ex]
			& & $ht\bar{t}$ & 7035 & 955 & 274 & $-$ & $-$ \\ [1.0ex]
			& & $Zt\bar{t}$ & 1483 & 220 & 54 & $-$ & $-$ \\ [1.0ex]
			\hline \hline 
			& BP1& $N_{S1}$ & 501 & 305 & 139 & 324 & 0.7046 \\ [1.0ex]
			\cline{2-5}
			Signals & & $N_{S2}$ & 717 & 417 & 185 & & \\ [1.0ex]
			\cline{2-8}
			& BP2 & $N_{S1}$ & 0 & 0 & 0 & 185 & 0.40238 \\ [1.0ex]
			\cline{2-5}
			& & $N_{S2}$ & 717 & 417 & 185 & & \\ [1.0ex]
			\cline{2-8}
			& BP3 & $N_{S1}$ & 0 & 0 & 0 & 234 & 0.5089 \\ [1.0ex]
			\cline{2-5}
			& & $N_{S2}$ & 372 & 296 & 234 & & \\ [1.0ex]
			\cline{2-8}
			& BP4 & $N_{S1}$ & 342 & 253 & 167 & 495 & 1.07637 \\ [1.0ex]
			\cline{2-5}
			& & $N_{S2}$ & 683 & 502 & 328 &  & \\ [1.0ex]
			\hline \hline
		\end{tabular}
		\caption{The cut flow table for the signals and backgrounds at 14 TeV for four benchmark points.}
		\label{table3}
	\end{table}	
	As both the signals have similar final states, in case of BP1 and BP4 $S$ is practically the collection of two different signal event numbers, $S=S_1+S_2$. In Table \ref{table4}, we present the case for 27 TeV LHC run with $3000$ fb$^{-1}$ integrated luminosity. Here we find that we obtain the maximum significance when we select only those events who pass the criteria of having $p_t^b >$ 230 GeV and $H_T >$ 680 GeV.	
	\begin{table}[H]
		\centering
		%\small
		\begin{tabular}{|c|c|c|c|c|c|c|c|}
			\hline \hline	
			& & & $n_b \ge 3$ & $p_{Tb1} > 230$ GeV & $H_T ~~> 680$ GeV & $S=N_{S1}+N_{S2}$ & $\mathcal{S}$ \\ [1.0ex]
			\hline \hline				
			& & $t\bar{t} + jets$ & 23648538 & 4122984 & 3326268 & $-$ & $-$ \\ [1.0ex]
			Backgrounds & & $b\bar{b}t\bar{t}$ & 197815 & 45933 & 38253 & $-$ & $-$ \\ [1.0ex]
			& & $ht\bar{t}$ & 29491 & 5831 & 4715 & $-$ & $-$ \\ [1.0ex]
			& & $Zt\bar{t}$ & 6489 & 1367 & 1034 & $-$ & $-$ \\ [1.0ex]
			\hline \hline 
			& BP1 & $N_{S1}$ & 7890 & 5427 & 5099 & 9379 & 5.10649 \\ [1.0ex]
			\cline{2-5}
			& & $N_{S2}$ & 7034 & 4584 & 4280 &  & \\ [1.0ex]
			\cline{2-8}
			Signals & BP2 & $N_{S1}$ & 0 & 0 & 0 & 4280 & 2.33087 \\ [1.0ex]
			\cline{2-5}
			& & $N_{S2}$ & 7034 & 4584 & 4280 &  & \\ [1.0ex]
			\cline{2-8}
			& BP3 & $N_{S1}$ & 0 & 0 & 0 & 2925 & 1.59305 \\ [1.0ex]
			\cline{2-5}
			& & $N_{S2}$ & 3543 & 2958 & 2925 &  & \\ [1.0ex]
			\cline{2-8}
			& BP4 & $N_{S1}$ & 3246 & 2562 & 2505 & 5538 & 3.01579 \\ [1.0ex]
			\cline{2-5}
			& & $N_{S2}$ & 3988 & 3110 & 3033 &  & \\ [1.0ex]
			\hline \hline
		\end{tabular}
		\caption{The cut flow table for the signals and backgrounds at 27 TeV for four benchmark points.}
		\label{table4}
	\end{table}	\normalsize
	
	From the above Tables, \ref{table3} and \ref{table4}, we observe that the significance for the case of 14 TeV HL-LHC run is much small which gets much better while considering the case for 27 TeV HL-LHC run. BP1 provides 0.7 significance ($\mathcal{S}$) for 14 TeV run whereas the significance increases to 5.1 for 27 TeV run at HL-LHC (Table \ref{table4}). But the results obtained using the cut-based approach does not make this method much useful. This motivates us to explore our results using multivariate analysis which we discuss in the following.

	At this point we would like to mention that the variable $H_T$ is highly correlated to the missing transverse energy (MET), $E_T \!\!\!\!\!\!/~~$. The correlation plots between $H_T$ and $E_T \!\!\!\!\!\!/~~$ are shown Fig. \ref{correlation_root} in the Appendix B for both signals. The distibutions of $E_T \!\!\!\!\!\!/~~$ and $H_T$ after applying the cut on $p_T$ of leading $b$-tagged jet are shown in Figs. \ref{aftercut}. In left figure, the distributions for signals and backgrounds are almost overlapped whereas in the right figure we can see a little seperation between signals and backgrounds. Instead of using $H_T$, one could use MET as a variable in the cut-based analysis. Using $E_T \!\!\!\!\!\!/~~$ as a variable we could find the signal significance $\mathcal{S}$ is maximum when $E_T \!\!\!\!\!\!/~~ >$ 20 GeV i.e. practically there is no extra cut on MET. The cut-flow table is shown in Table \ref{table6} in the Appendix. This behaviour along with the above mentioned plots made us choose $H_T$ as a better signal-background discriminating variable in the cut-based analysis even if the significance is smaller in this case.
	
	\subsection{Multivariate Analysis}
	
	In this section, we mainly concentrate on the results we obtain using Boosted Decision Tree (BDT) algorithm. This part of analysis has been performed in a \texttt{TMVA} framework \cite{tmva}. 
	Decision trees are mainly classifiers who generally classify the signal and background-like events. A suitable variable is chosen and application of a proper cut on this variable separates the signal from the background as best as it can. One can choose a number of variables and train the signal and background sample events. Modification of the weights corresponding to the sample events creates new \emph{boosted} decision trees. After training and testing of the signal and background-like events, this method of analysis excels the generic cut-based analysis by performing a much better discrimination between signal events and background events.
	
	To perform the BDT analysis we have considered 11 variables, providing the best possible signal significance, are as the following.
	
	\begin{itemize}
		\item The transverse momentum of leading b-tagged jet, $p_T^{b1}$. $b_1$ denotes that b-tagged jet which has the highest transverse momentum.
		\item The missing transverse energy, $E_T \!\!\!\!\!\!/~~$.
		\item The transverse momentum of leading lepton, $p_T^{l1}$. Again, $l_1$ denotes the lepton which has the highest transverse momentum.
		\item $\Delta\eta^{bibj}$ between the b-tagged jets. $i$ and $j$ stand for the numbers 1, 2 and 3. $b_1, b_2, b_3$ denote three $p_T$ ordered b-tagged jets in descending order respectively. This variable will provide the difference between pseudorapidity of two b-tagged jets. For example, $\Delta\eta^{b1b2}$ implies the difference between $\eta$s of leading and sub-leading b-jets.
		\item $\Delta\phi^{bibj}$ between the b-tagged jets. This variable implies difference between the azimuthal angles corresponding to two b-tagged jets.
		\item $\Delta\eta^{l1l2}$ between the leading and sub-leading lepton.
		\item $\Delta\phi^{l1l2}$ between the leading and sub-leading lepton.
	\end{itemize}
	
	\begin{table}[H]
		\centering
		\begin{tabular}{|c|c|c|}
			\hline \hline
			Rank & Variable Importance & Variable Importance \\ [1.0ex]
			& (14 TeV) & (27 TeV) \\ [1.0ex]
			\hline \hline
			1 & $E_T \!\!\!\!\!\!/~~$ & $E_T \!\!\!\!\!\!/~~$ \\ [1.0ex]
			2 & $\Delta\phi^{b1b2}$ & $\Delta\phi^{b1b3}$ \\ [1.0ex]
			3 & $\Delta\phi^{l1l2}$ & $\Delta\phi^{b1b2}$ \\ [1.0ex]
			4 & $\Delta\phi^{b1b3}$ & $\Delta\phi^{l1l2}$ \\ [1.0ex]
			5 & $\Delta\phi^{b2b3}$ & $\Delta\eta^{l1l2}$ \\ [1.0ex]
			6 & $\Delta\eta^{l1l2}$ & $p_T^{b1}$ \\ [1.0ex]
			7 & $p_T^{b1}$ & $\Delta\phi^{b2b3}$ \\ [1.0ex]
			8 & $\Delta\eta^{b1b3}$ & $\Delta\eta^{b1b2}$ \\ [1.0ex]
			9 & $\Delta\eta^{b2b3}$ & $\Delta\eta^{b1b3}$ \\ [1.0ex]
			10 & $\Delta\eta^{b1b2}$ & $\Delta\eta^{b2b3}$ \\ [1.0ex]
			11 & $p_T^{l1}$ & $p_T^{l1}$ \\ [1.0ex]
			\hline \hline
		\end{tabular}
		\caption{Importance of the variables used in the BDT analysis}
		\label{rank}
	\end{table}
	
	The Table \ref{rank} shows the ranks of the above variables according to their relevance for both 14 and 27 TeV signal-background study. The rank of the variables are determined depending on how many times a variable is used to split decision tree nodes. Here in both cases of 14 and 27 TeV run, $E_T \!\!\!\!\!\!/~~$ has been the most important variable. In our study, the important parameters for a BDT analysis have been set as follows. We have set the number of Trees to be 850 with maximum depth 3 and the boost type as AdaBoost. 
	
	The normalized distributions of the above variables are shown in Fig. \ref{bdt_dist}. The blue-shaded (red-dashed) distributions are for the signal (background). It is to be mentioned that while doing this analysis, all the four backgrounds have been taken into consideration despitethe fact that $t\bar{t} + jets$ production is the most dominating one. 
	
	\begin{figure}[H]
		\begin{subfigure}{0.49\textwidth}
			\begin{center}
				\includegraphics[width=7.5cm, height=5cm]{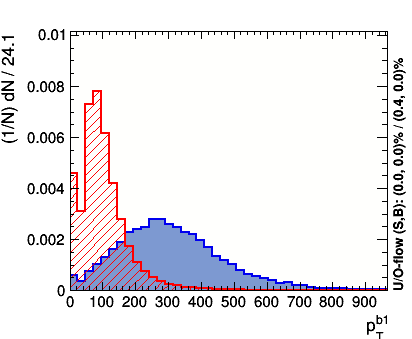}
			\end{center}
			\caption{}
		\end{subfigure}
		\begin{subfigure}{0.49\textwidth}
			\begin{center}
				\includegraphics[width=7.5cm, height=5cm]{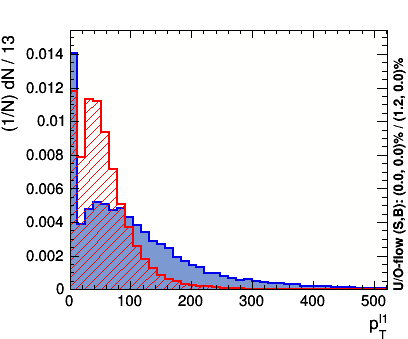}
			\end{center}
			\caption{}
		\end{subfigure}
		\begin{subfigure}{0.49\textwidth}
			\begin{center}
				\includegraphics[width=7.5cm, height=5cm]{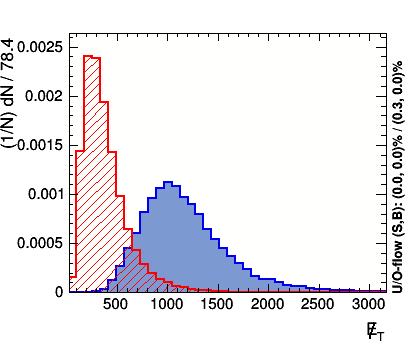}
			\end{center}
			\caption{}
		\end{subfigure}
		\begin{subfigure}{0.49\textwidth}
			\begin{center}
				\includegraphics[width=7.5cm, height=5cm]{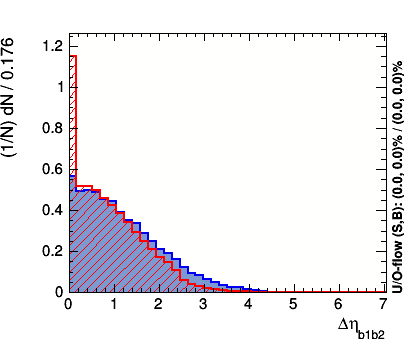}
			\end{center}
			\caption{}
		\end{subfigure}
		\begin{subfigure}{0.49\textwidth}
			\begin{center}
				\includegraphics[width=7.5cm, height=5cm]{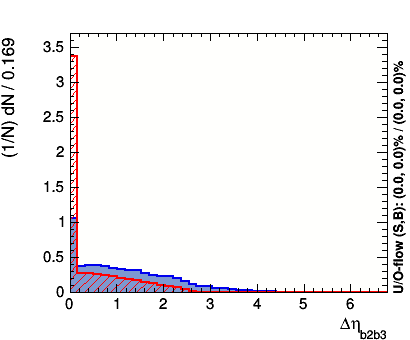}
			\end{center}
			\caption{}
		\end{subfigure}
		\begin{subfigure}{0.49\textwidth}
			\begin{center}
				\includegraphics[width=7.5cm, height=5cm]{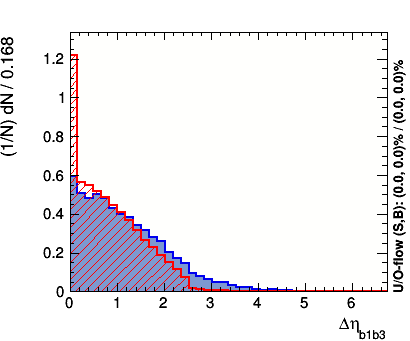}
			\end{center}
			\caption{}
		\end{subfigure}
	\end{figure}
		
	\begin{figure}[H]\ContinuedFloat	
		\begin{subfigure}{0.49\textwidth}
			\begin{center}
				\includegraphics[width=7cm, height=5cm]{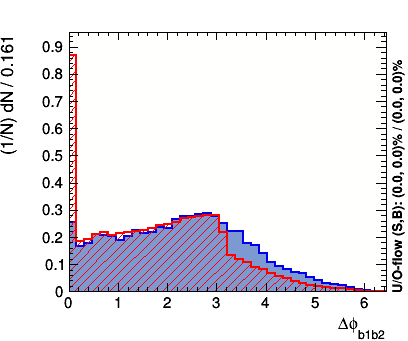}
			\end{center}
			\caption{}
		\end{subfigure}
		\begin{subfigure}{0.49\textwidth}
			\begin{center}
				\includegraphics[width=7cm, height=5cm]{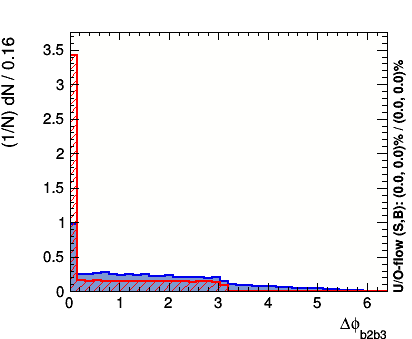}
			\end{center}
			\caption{}
		\end{subfigure}
		\begin{subfigure}{0.33\textwidth}
			\begin{center}
				\includegraphics[width=5.7cm, height=5cm]{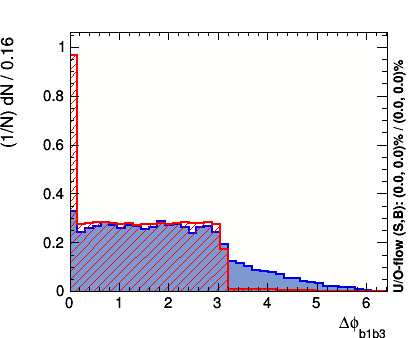}
			\end{center}
			\caption{}
		\end{subfigure}
		\begin{subfigure}{0.33\textwidth}
			\begin{center}
				\includegraphics[width=5.7cm, height=5cm]{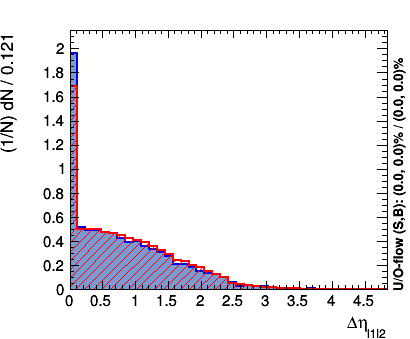}
			\end{center}
			\caption{}
		\end{subfigure}
		\begin{subfigure}{0.33\textwidth}
			\begin{center}
				\includegraphics[width=5.7cm, height=5cm]{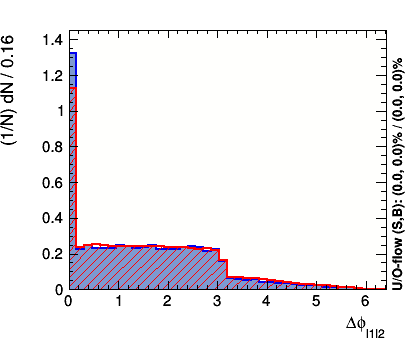}
			\end{center}
			\caption{}
		\end{subfigure}
		
		\caption{Distribution plots of the variables we have accounted for multivariate analysis for BP1, at 14 TeV HL-LHC run.}
		\label{bdt_dist}	
	\end{figure}

	\begin{figure}[H]
		\begin{subfigure}{0.49\textwidth}
			\begin{center}
				\includegraphics[width=8.7cm, height=6.8cm]{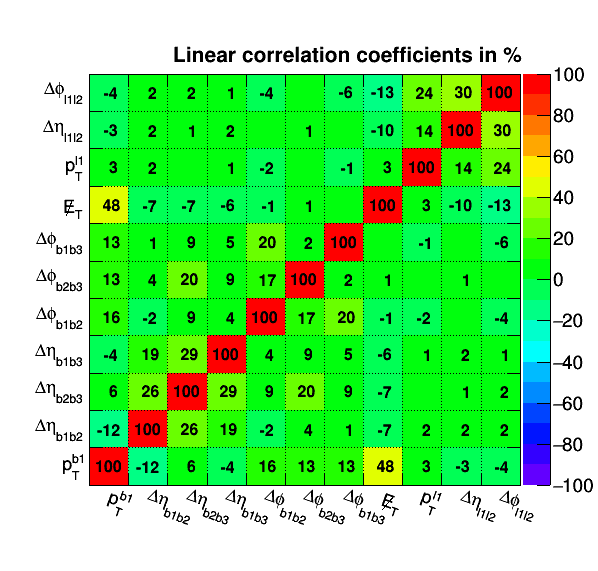}
			\end{center}
			\caption{For Signal}
		\end{subfigure}
		\begin{subfigure}{0.49\textwidth}
			\begin{center}
				\includegraphics[width=8.7cm, height=6.8cm]{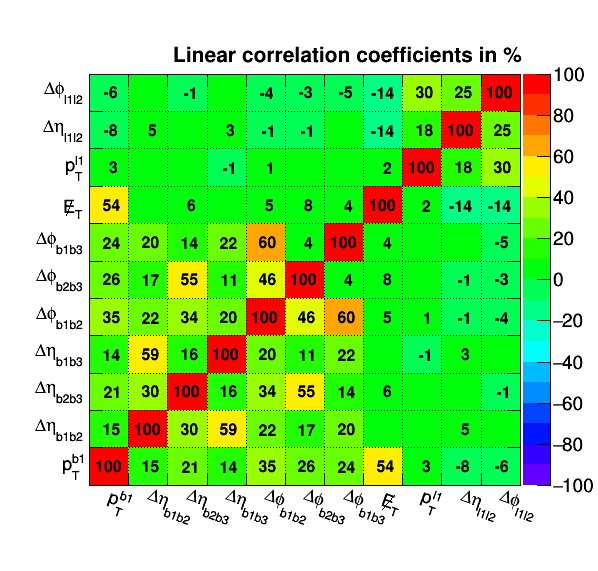}
			\end{center}
			\caption{For Background}
		\end{subfigure}
		\caption{The linear correlations between the variables considered for the multivariate analysis are shown here in form of percentage for signal (a) as well as background (b) for benchmark point BP1 at 14 TeV HL-LHC run. The negetive sign implies that the two corresponding variables are anti-correlated.}
		\label{correlation}	
	\end{figure}
	The linear correlation matrix for the variables of our choice is shown below in Fig. \ref{correlation} for only benchmark point BP1. The correlations between any two variables are presented in \% in this figure. One can see, in most of the cases, the variables are not correlated in a significant way.
	
	\begin{figure}[H]
		\begin{subfigure}{0.49\textwidth}
			\begin{center}
				\includegraphics[width=8cm, height=6cm]{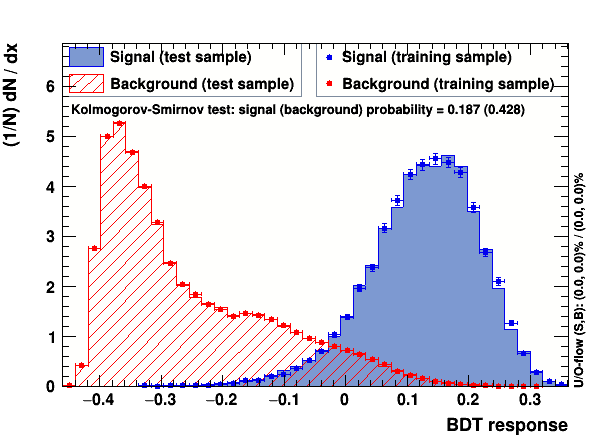}
			\end{center}
			\caption{For BP1 (14 TeV)}
		\end{subfigure}
		\begin{subfigure}{0.49\textwidth}
			\begin{center}
				\includegraphics[width=8cm, height=6cm]{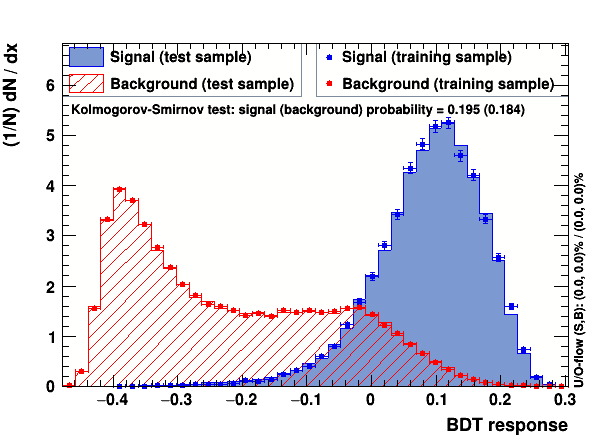}
			\end{center}
			\caption{For BP1 (27 TeV)}
		\end{subfigure}
		\caption{The result of Kolmogorov-Smirnov test for BP1 at 14 TeV (a) and 27 TeV (b) LHC run with integrated luminosity 3000 fb$^{-1}$ respectively.}
		\label{overtrain}	
	\end{figure}
	
	The signal and background events have been trained for each four benchmark points. A partial overtraining might be quite possible for boosted decision tree algorithm which must be avoided. It can be tested comparing the performance of training and testing samples. We have ensured that the effect of ovetraining of signal and background is minimal for our cases by Kolmogorov-Smirnov (KS) test. In general, KS score must be $\sim 0.1$. It may be greater than 0.01 if this value remains fixed over changing the statistics of the signal and background events. In Fig. \ref{overtrain}, one can see the value of the KS probability is $\sim 0.187 (0.428)$ and $\sim 0.195 (0.184)$ for signal (background) for BP1 at 14 TeV and 27 TeV HL-LHC run.
	
	\begin{figure}[H]
		\begin{subfigure}{0.49\textwidth}
			\begin{center}
				\includegraphics[width=8cm, height=6cm]{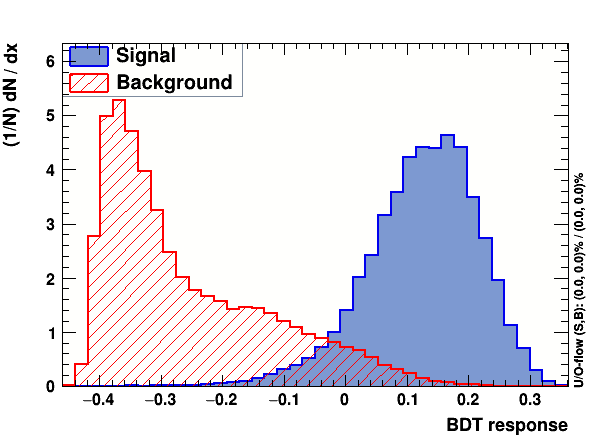}
			\end{center}
			\caption{For BP1 (14 TeV)}
		\end{subfigure}
		\begin{subfigure}{0.49\textwidth}
			\begin{center}
				\includegraphics[width=8cm, height=6cm]{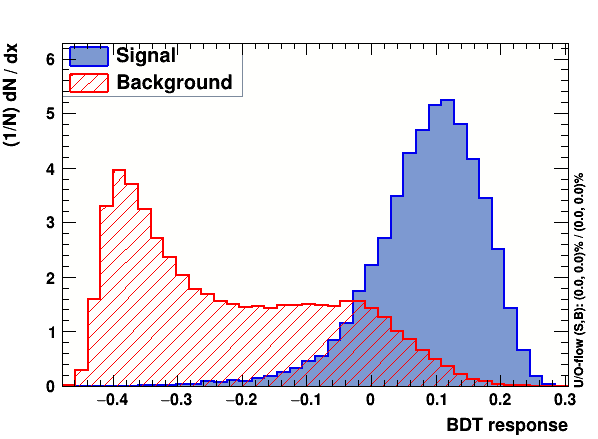}
			\end{center}
			\caption{For BP1 (27 TeV)}
		\end{subfigure}
		
		\caption{The BDT response of the signal and backgrounds for 14 and 27 TeV HL-LHC run for the BP1 respectively.}
		\label{sig_bkg_14}	
	\end{figure}
	
	Half of the signal and background events have been used for training and the other half of the same sample is used for testing. After a successful training and testing of the signal and background samples, the BDT algorithm has made the results for 14 as well a 27 TeV HL-LHC run better compared to cut-based analysis. The TMVA response of the classification has shown an good discrimination between signal and background which is shown in Fig. \ref{sig_bkg_14} for BP1 benchmark point at 14 as well as 27 TeV HL-LHC run.
	
\begin{figure}[H]
		\begin{subfigure}{0.49\textwidth}
			\begin{center}
				\includegraphics[width=8.5cm, height=6cm]{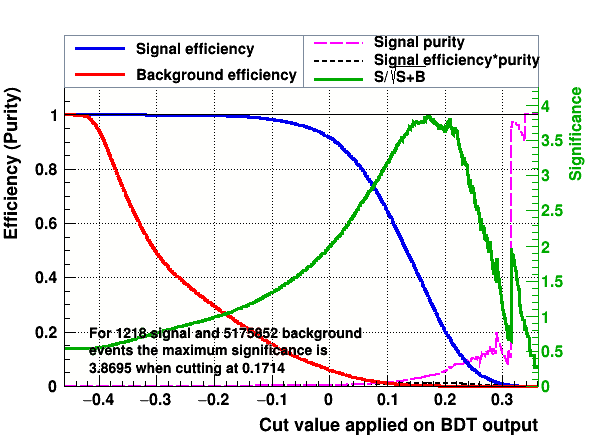}
			\end{center}
			\caption{For BP1 (14 TeV)}
		\end{subfigure}
		\begin{subfigure}{0.49\textwidth}
			\begin{center}
				\includegraphics[width=8.5cm, height=6cm]{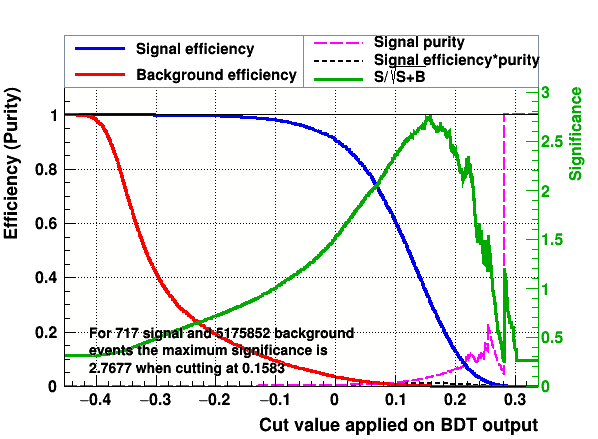}
			\end{center}
			\caption{For BP2 (14 TeV)}
		\end{subfigure}
		\begin{subfigure}{0.49\textwidth}
			\begin{center}
				\includegraphics[width=8.5cm, height=6cm]{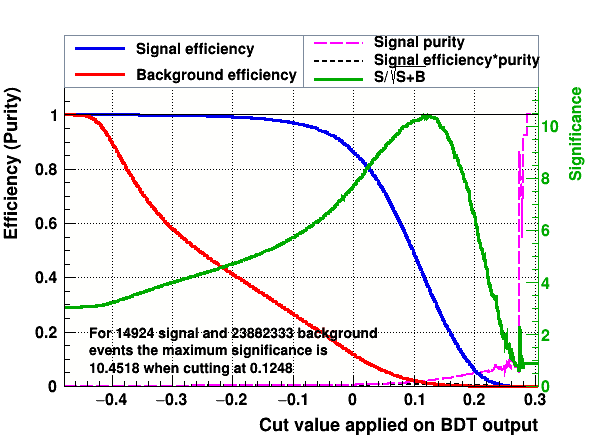}
			\end{center}
			\caption{For BP1 (27 TeV)}
		\end{subfigure}
		\begin{subfigure}{0.49\textwidth}
			\begin{center}
				\includegraphics[width=8.5cm, height=6cm]{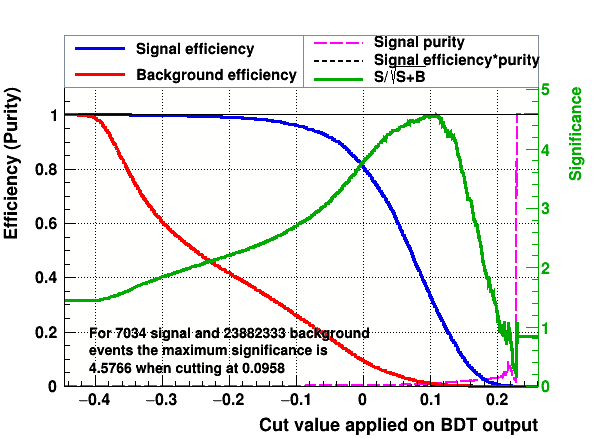}
			\end{center}
			\caption{For BP2 (27 TeV)}
		\end{subfigure}
		
		\caption{The signal and background efficiency and significance for 14 and 27 TeV HL-LHC run for BP1 and BP2 respectively.}
		\label{significancetmva}	
	\end{figure}

	The significance we calculate using the expression shown in Eq. \ref{significance} has improved in a significant amount compared to cut-based scenario which is explained in Table \ref{table5} for both 14 and 27 TeV run of LHC for all of the benchmark points respectively. In Fig. \ref{significancetmva} the signal efficiency, background efficiency and signal significance have been presented for two benchmark points BP1 and BP2 at 14 and 27 TeV HL-LHC run where the results for BP2 solely corresponds to probing a charged Higgs at the hadron collider.
	
	The significances obtained from the BDT analysis for each case have been given below in a form of a table (see Table \ref{table5}). The significance obtained for BP1 is $\sim 3.87$ which is much better compared to the significance achieved in case of cut-based analysis as one can expect. Similar increments are observed for the other benchmark points BP2, BP3, BP4 at 14 TeV run. The results obtained for 27 TeV HL-LHC run are more encouraging. From the results for BP2 and BP3, in 32121 model one can hope to probe a charged Higgs of mass 750 GeV and 1.2 TeV with 2.77 $\sigma$ (4.58 $\sigma$) and 1.38 $\sigma$ (3.66 $\sigma$) significance respectively at 14 (27) TeV HL-LHC run (see Figs. \ref{significancetmva} (b), \ref{significancetmva} (d) for BP2).
	
	\begin{table}[t]
		\centering
		\begin{tabular}{|c|c|c|}
			\hline \hline
			& $\mathcal{S}$ (14 TeV) & $\mathcal{S}$ (27 TeV) \\ [1.0ex]
			\hline 
			BP1 & 3.8695 & 10.4518 \\ [1.0ex]
			BP2 & 2.7677 & 4.5766 \\ [1.0ex]
			BP3 & 1.3838 & 3.6577 \\ [1.0ex]
			BP4 & 2.5171 & 6.5159 \\
			\hline\hline
		\end{tabular}
		\caption{The significances obtained with multivariate analysis for each benchmark point at 14 and 27 TeV LHC run}
		\label{table5}
	\end{table}	
	
	Before we close this section, want to mention an important point regarding our analysis. Boosted Decision Tree (BDT) method works very differently that of the cut-based method. In case of cut-based analysis we have used selection criterion on chosen kinematic variables one after another.  But in BDT, different cuts on all the variables are applied to classify the events to check whether they are signal-like events or background-like events. BDT, in turn, optimise the selection criterion collectively such that the signal background separation becomes optimal. 
	So it is not very easy to comment why BDT has found MET as the most effective variable which is not the case for the cut-based method, as  we are unable to apply different cuts on all variables at the same time and check the signal significances. Moreover, we observe that $H_T$ and MET are strongly correlated (with correlation coefficient 0.4). We have redone the whole MVA analysis considering $H_T$ instead of MET. This basically lowers the signal significance compared to the previous case. For example for 14 TeV HL LHC run, signal significance for BP1 has reduced to 9.87 from 10.45 when using $H_T$ as one of the 11 variables. The rank of $H_T$ is also much lower compared to MET. 
		
	Furthermore in this analysis, while performing BDT we could find that some of the variables are not much correlated. We have performed the same analysis with less number of variables. More specifically we have omitted the least important variable $p_T^{l1}$ and repeated the same analysis again. We have obtained smaller significance $\mathcal{S}$ for each benchmark point. So as a consequence, we have decided to keep all 11 variables we have used in this study.
	
	\section{Conclusions}
	\label{sec4}	
	
	To summarise, we have investigated the possible collider signatures of heavy Higgs boson arising in an extension of the Standard Model with local gauge symmetry $SU(3)_C \otimes SU(2)_L \otimes U(1)_L \otimes SU(2)_R \otimes U(1))_R$. Breaking of  Left-Right symmetry  from $SU(3)_C \otimes SU(2)_L \otimes U(1)_L \otimes SU(2)_R \otimes U(1))_R$ down to $SU(3)_C \otimes SU(2)_L \otimes U(1)_Y$ requires  several other Higgs multiplet apart from the SM-like one.  After symmetry breaking, six heavy neutral scalars (4 CP-even 
	and 2 CP-odd) and two charged scalars remain in the physical spectrum. One such scalars can be identified with the SM like Higgs boson with mass close to 125 GeV. 
	
	The full \textbf{27}-plet of fermions arising in $E_6$, (from whose breaking in more than one steps, results in to the gauge group of our interest) have been considered in the present analysis. 
	
	The gauge sector of the  model contains five gauge couplings whose values have been fixed following the pattern of Left-Right symmetry breaking. Apart from the SM gauge bosons this model contains $W'$, $Z'$ and $A'$ gauge bosons where $A'$ is the hallmark of the extra $U(1)$ gauge symmetry. Experimental limits on the gauge boson masses will in turn set limits on the Higgs boson vacuum expectation values. 
	
	In this article we have mainly set our focus on three  heavy Higgs bosons, a  neutral CP-even Higgs field, $h_2^0$ and its CP-odd partner $\xi_2^0$ along with  a singly charged scalar, $H_1^\pm$. The neutral scalars  have similar masses and couplings as both of them  arise from the Higgs bi-doublet $\Phi_B$. $h_2^0 ~(\xi_2^0)$ dominantly decays to $b\bar{b}$ until the decay channel to $H_1^\pm W^\mp$ is kinematically accessible. $H_1^+$ dominantly decays to $t\bar{b}$ until the decay channel to $h_2^0 ~(\xi_2^0) W^+$ is kinematically allowed. We have used these informations on BSM decay channels while discussing about the signatures of these scalars at the LHC. For $h_2^0 ~(\xi_2^0)$ we have mainly chosen the dominant production mechanism of this scalar which is in our case the associated Higgs production. The production cross-section of $h_2^0 ~(\xi_2^0)$ in association of $b\bar{b}$ is 0.3 (3) pb at 14 (27) TeV LHC run for 1 TeV mass. Whereas $H_1^\pm$ has been produced in association with $tb$. Production cross-section of $H_1^\pm$ in association of $tb$ is 0.04 (0.35) pb at 14 (27) TeV LHC run for 1 TeV scalar mass.
	
	We have performed a detailed signal-background analysis of two of the heavy Higgs bosons, $h_2^0 ~(\xi_2^0)$ and $H_1^\pm$. The associated production of both of them give rise to similar final states with three or more than three b-tagged jets, more than one charged leptons and missing transverse energy. The dominant background will arise from $t\bar{t}$ production with jets. The other background events will arise from $b\bar{b}t\bar{t}$, $ht\bar{t}$, $Zt\bar{t}$ productions. Depending on the masses and decay properties of the heavy neutral and charged scalars in our model, we choose four benchmark points (BP) to perform our analysis. To begin with, we have presented our results using cut-based analysis for four benchmark points. We have applied a series of cuts on some chosen suitable variables like transverse momentum ($p_T$) of leading b-tagged jet and the scalar sum of $p_T$ of all jets ($H_T$). For BP1, at 14 (27) TeV the signal to background ratio is 0.7 (5.1) whereas for other benchmark points this is somewhat lower except the case for BP4 at 14 TeV run ($\mathcal{S} \sim$ 1.1). In order to distinguish signal events from background-like events more accurately we have used a better algorithm used in multivariate analysis where we have chosen the BDT method. With this mechanism, as per our expectation,  a better significance could be achieved for all of the benchmark points. For BP1, at 14 (27) TeV the significance is 3.87 (10.45) which clearly shows a better signal-background discrimination. With the results we obtained, one can hope to probe a heavy charged Higgs of a mass 750 GeV in the 32121 model, with $2.77 \sigma$ ($4.58 \sigma$) significance at a 14 (27) TeV LHC run with 3000 fb$^{-1}$ integrated luminosity.
	
	\vspace{5mm}
	{\bf Acknowledgement:} SB acknowledges financial support from DST, Ministry of Science and Technology, Government of India in the form of an INSPIRE-Senior Research Fellowship. SB acknowledges Prof. Anindya Datta for his valuable suggestions throughout the analysis. SB also acknowledges Gourab Saha and Nivedita Ghosh for their help in dealing with some technical issues. SB is thankful to Prof. Partha Konar for the insightful discussions. SB also acknowledges Prof. Satyaki Bhattacharya for his useful advice.

	\section{Appendix:}
	
	\appendix
	
	\section{List of the Feynman rules used for this analysis:}
	
	\begin{table}[H]
		\centering
		\begin{tabular}{|c|c|}
			\hline \hline
			Interactions & Couplings \\ [1.0ex]
			\hline \hline	
			$h_2^0 b \bar{b}$ & $i \dfrac{1}{\sqrt{2}}y_t$  \\ [1.0ex]
			\hline
			$\xi_2^0 b \bar{b}$ & $i \dfrac{1}{\sqrt{2}}y_t \gamma_5$  \\ [1.0ex]
			\hline
			$H_1^+ \bar{t} b$ & $i \sqrt{1-\dfrac{k_1^2}{v_R^2}} (y_t  + y_b) V_{tb}$ \\ [1.0ex]
			\hline
			$h_2^0 H_1^+ W^-$ & $i \dfrac{g_{2L}}{2} \sqrt{1-\dfrac{k_1^2}{v_R^2}}$ \\ [1.0ex]
			\hline \hline
		\end{tabular}
		\caption{Feynman rules used in this analysis.}
		\label{coupling}
	\end{table}

	\section{Additional plots and tables related to collider phenomenology}
	
	The correlation between $H_T$ and MET is shown in the below diagrams for Signal 1 and Signal 2 respectively.
	\begin{figure}[H]
		\centering
		\includegraphics[width=8.5cm, height=7cm]{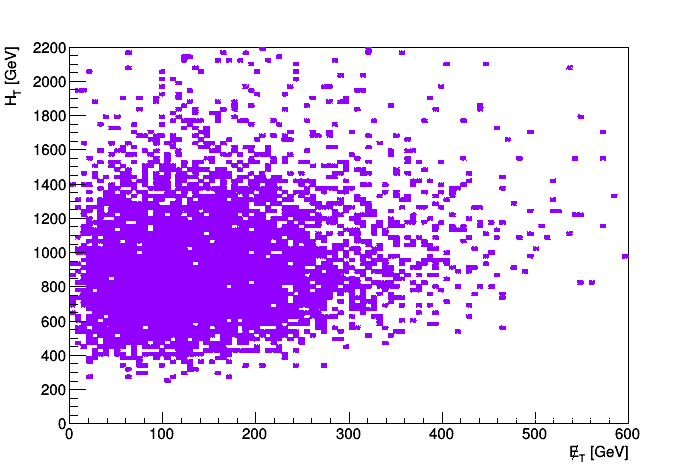}
		\includegraphics[width=8.5cm, height=7cm]{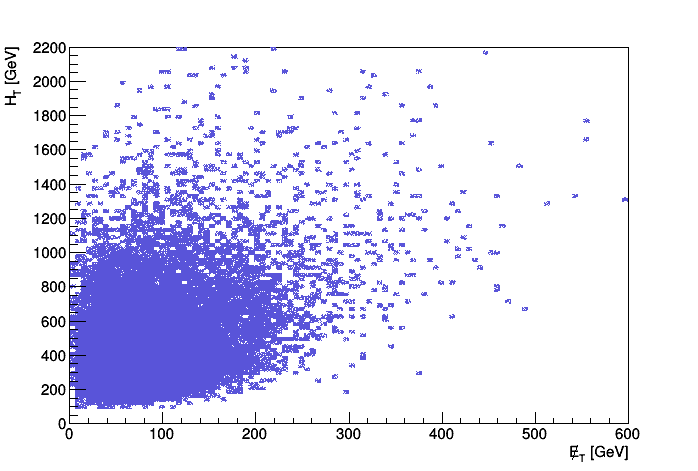}
		\caption{The correlaion plots between MET and HT for Signal 1 (left) and $t\bar{t}$+jets background (right) at 14 TeV LHC run}
		\label{correlation_root}
	\end{figure}

	The below diagrams represent the distribution of MET and $H_T$ after we apply the cut on the transverse momentum of leading $b$-tagged jet.
	\begin{figure}[H]
		\centering
		\includegraphics[width=8.5cm, height=7cm]{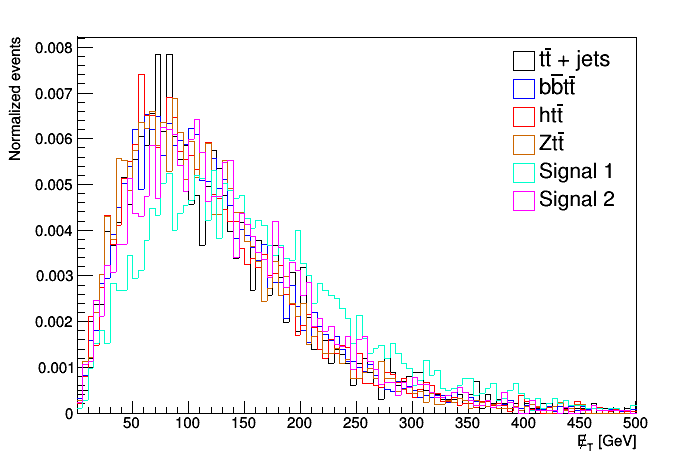}
		\includegraphics[width=8.5cm, height=7cm]{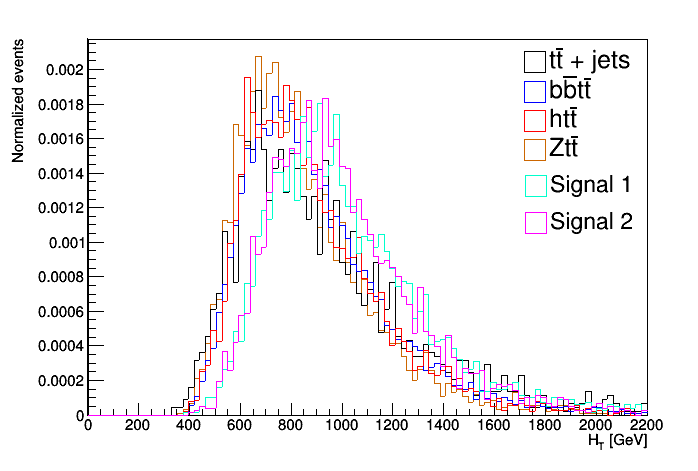}
		\caption{Distribution of MET (left) and $H_T$ (right) at 14 TeV LHC run after applying cut on $p_{T}$ ($p_{T} > 240$ GeV)}.
		\label{aftercut}
	\end{figure}	
	Unlike the left diagram, the right diagram shows a little seperation between backgrounds and signals. The cut-flow table using MET as a variable is shown below.
	
	\begin{table}[H]
		\centering
		%\small
		\begin{tabular}{|c|c|c|c|c|c|c|c|}
			\hline \hline	
			& & & $N_b \ge 3$ & $p_{Tb1} > 240$ GeV & $MET ~~> 20$ GeV & $S=N_{S1}+N_{S2}$ & $\mathcal{S}$ \\ [1.0ex]
			\hline \hline				
			& & $t\bar{t} + jets$ & 5127095 & 624321 & 610111 & $-$ & $-$ \\ [1.0ex]
			Backgrounds & & $b\bar{b}t\bar{t}$ & 40239 & 6228 & 6089 & $-$ & $-$ \\ [1.0ex]
			& & $ht\bar{t}$ & 7035 & 955 & 933 & $-$ & $-$ \\ [1.0ex]
			& & $Zt\bar{t}$ & 1483 & 220 & 215 & $-$ & $-$ \\ [1.0ex]
			\hline \hline 
			& BP1& $N_{S1}$ & 501 & 305 & 301 & 708 & 0.9 \\ [1.0ex]
			\cline{2-5}
			Signals & & $N_{S2}$ & 717 & 417 & 407 & & \\ [1.0ex]
			\cline{2-8}
			& BP2 & $N_{S1}$ & 0 & 0 & 0 & 407 & 0.5179 \\ [1.0ex]
			\cline{2-5}
			& & $N_{S2}$ & 717 & 417 & 407 & & \\ [1.0ex]
			\cline{2-8}
			& BP3 & $N_{S1}$ & 0 & 0 & 0 & 291 & 0.3703 \\ [1.0ex]
			\cline{2-5}
			& & $N_{S2}$ & 372 & 296 & 291 & & \\ [1.0ex]
			\cline{2-8}
			& BP4 & $N_{S1}$ & 342 & 253 & 250 & 744 & 0.9467 \\ [1.0ex]
			\cline{2-5}
			& & $N_{S2}$ & 683 & 502 & 494 &  & \\ [1.0ex]
			\hline \hline
		\end{tabular}
		\caption{The cut flow table for the signals and backgrounds at 14 TeV for four benchmark points.}
		\label{table6}
	\end{table}


\begin{thebibliography}{widestlabel}
		
		
		\bibitem{higgs_atlas} ATLAS collaboration, G. Aad \emph{et al.}, \emph{Observation of a new particle in the search for the Standard Model Higgs boson with the ATLAS detector at the LHC}, \href{https://doi.org/10.1016/j.physletb.2012.08.020}{Phys. Lett. B716 (2012) 1-29}, arXiv: [\href{https://arxiv.org/pdf/1207.7214.pdf}{1207.7214}].
		
		%%%%%			
		
		\bibitem{higgs_cms} CMS collaboration, S. Chatrchyan \emph{et al.}, \emph{Observation of a New Boson at a Mass of 125 GeV with the CMS Experiment at the LHC}, \href{https://doi.org/ 	10.1016/j.physletb.2012.08.021}{Phys. Lett. B 716 (2012) 30}, arXiv: [\href{https://arxiv.org/pdf/1207.7235.pdf}{1207.7235}].
		
		%%%%%		
		
		\bibitem{Higgs-review} M. M\"{u}hlleitner, M. O. P. Sampaio, R. Santos and J. wittbrodt, \emph{Phenomenological comparison of models with extended Higgs sectors}, \href{https://doi.org/10.1007/JHEP08(2017)132}{JHEP08 (2017) 132}, arXiv: [\href{https://arxiv.org/pdf/1703.07750.pdf}{1703.07750}]; 
		\hspace{5mm}
		J. Steggemann, \emph{Extended Scalar Sectors}, \href{https://doi.org/10.1146/annurev-nucl-032620-043846}{Annu. Rev. Nucl. Part. Sci. 2020. 70:197–223} and references therein.
		
		%%%%%			
		
		\bibitem{higgs-precision1} J. Alison \emph{et al.}, \emph{Higgs boson potential at colliders: status and perspectives}, \href{https://doi.org/10.1016/j.revip.2020.100045}{Review in Physics (2020) 100045}, arXiv: [\href{https://arxiv.org/pdf/1910.00012.pdf}{1910.00012}]. 
		
		%%%%%
		
		\bibitem{higgs-precision2} G. Heinrich, \emph{Collider Physics at the Precision Frontier}, \href{https://doi.org/10.1016/j.physrep.2021.03.006}{Physics Reports, Volume 922, 2021, Pages 1-69}, arXiv: [\href{https://arxiv.org/pdf/2009.00516.pdf}{2009.00516}].
		
		%%%%%
		
		\bibitem{dihiggs} F. Arco, S. Heinemeyer, M. Mhlleitner and K. Radchenko, \emph{Sensitivity to Triple Higgs Couplings via Di-Higgs Production in the 2HDM at the (HL-)LHC}, \href{https://doi.org/10.1140/epjc/s10052-023-12193-4}{Eur.Phys.J.C 83 (2023) 11, 1019},
		arXiv: [\href{https://arxiv.org/pdf/2212.11242.pdf}{2212.11242}];
		\hspace{5mm}
		H. Abouabid, A. Arhrib, D. Azevedo, J.E. Falaki, P.M. Ferreira, M. Mhlleitner et al., \emph{Benchmarking di-Higgs production in various extended Higgs sector models}, \href{https://doi.org/10.1007/JHEP09(2022)011}{JHEP 09 (2022) 011}, arXiv: [\href{https://arxiv.org/pdf/2112.12515.pdf}{2112.12515}].
		
		%%%%%			
		
		\bibitem{singlet-DM} C. E. Yaguna, \emph{The singlet scalar as FIMP dark matter}, \href{https://doi.org/10.1007/JHEP08(2011)060}{ 	JHEP08(2011)060}, arXiv: [\href{https://arxiv.org/pdf/1105.1654.pdf}{1105.1654}]; 
		\hspace{5mm}
		R. Campbell, S. Godfrey, H. E. Logan and A. Poulin, \emph{Real singlet scalar dark matter extension of the Georgi-Machacek model}, [\href{https://doi.org/10.1103/PhysRevD.95.016005}{Phys. Rev. D 95, 016005}]; 
		\hspace{5mm}
		The GAMBIT Collaboration, \emph{Status of the scalar singlet dark matter model}, [\href{https://doi.org/10.1140/epjc/s10052-017-5113-1}{Eur. Phys. J. C (2017) 77:568}]; 
		\hspace{5mm}
		P. Das, M. K. Das and N. Khan, \emph{A new feasible dark matter region in the singlet scalar scotogenic model}, [\href{https://doi.org/10.1016/j.nuclphysb.2021.115307}{Nuclear Physics B, Vol. 964, 115307}];
		
		%%%%%
		
		\bibitem{numass} E. Ma and O. Popov, \emph{Pathways to naturally small Dirac neutrino masses}, \href{https://doi.org/10.1016/j.physletb.2016.11.027}{Phys. Lett. B.2016.11.027}, arXiv: [\href{https://arxiv.org/pdf/1609.02538.pdf}{1609.02538}].
		
		%%%%%			
		
		\bibitem{triplet-neutrinomass} R. N. Mohapatra and P. B. Pal, \emph{Massive neutrinos in physics and astrophysics}, \href{https://doi.org/10.1142/5024}{World Sci. Lect. Notes Phys.72, 1 (2004)}; 
		\hspace{5mm}
		N. G. Deshpande, J. F. Gunion, B. Kayser, and F. Olness, \emph{Left-right-symmetric electroweak models with triplet Higgs field}, \href{https://doi.org/10.1103/PhysRevD.44.837}{Phys. Rev. D 44, 837};
		\hspace{5mm}
		E. Ma and U. Sarkar, \emph{Neutrino Masses and Leptogenesis with Heavy Higgs Triplets}, \href{https://doi.org/10.1103/PhysRevLett.80.5716}{Phys. Rev. Lett. 80 (1998) 5716-5719}, arXiv: [\href{https://arxiv.org/pdf/hep-ph/9802445.pdf}{hep-ph/9802445}].
		
		%%%%%
		
		\bibitem{Nu_mass1} C. Hati, S. Patra, P. Pritimita and U. Sarkar, \emph{Neutrino Masses and Leptogenesis in Left-Right Symmetric Models: A Review From a Model Building Perspective}, \href{https://doi.org/10.3389/fphy.2018.00019}{Front. Phys., 06 March 2018}. 
		
		%%%%%			
		
		\bibitem{2hdm} A. Vicente, \emph{Higgs Lepton Flavor Violating Decays in Two Higgs Doublet Models}, \href{https://doi.org/10.3389/fphy.2019.00174}{Front. Phys., fphy.2019.00174}; 
		\hspace{5mm}
		D. Das, P M. Ferreira, A. P. Morais, I. Padilla-Gay, R. Pasechnik and J. P. Rodrigues, \emph{A three Higgs doublet model with symmetry-suppressed flavour changing neutral currents}, \href{https://doi.org/10.1007/JHEP11(2021)079}{JHEP 11 (2021) 079}, arXiv: [\href{https://arxiv.org/pdf/2106.06425.pdf}{2106.06425}]; 
		\hspace{5mm}
		S. Iguro, Y. Muramatsu, Y. Omura and Y. Shigekami, \emph{Flavor physics in the multi-Higgs doublet models induced by the left-right symmetry}, \href{https://doi.org/10.1007/JHEP11(2018)046}{JHEP11 (2018) 046}, arXiv: [\href{https://arxiv.org/pdf/1804.07478.pdf}{1804.07478}].	
		
		%%%%%
		
		\bibitem{32121} S. Bhattacharyya and A. Datta, \emph{Phenomenology of an $E_6$ inspired extension of Standard Model: Higgs sector}, \href{https://doi.org/10.1103/PhysRevD.105.075021}{Phys. Rev. D 105, 075021}, arXiv: [\href{https://arxiv.org/pdf/2109.08524.pdf}{2109.08524}].
		
		%%%%%		
		
		\bibitem{E6} Y. Achiman and B. Stech, \emph{Quark-Lepton Symmetry and mass scales in an E6 unified gauge model}, \href{https://doi.org/10.1016/0370-2693(78)90584-1}{Physics Letters B, 77(4-5), 389-393}; 
		\hspace{5mm}
		Q. Shafi, \emph{E6 as a unifying gauge symmetry}, \href{https://doi.org/10.1016/0370-2693(78)90248-4}{Physics Letters B, 79(3), 301-303}; 
		\hspace{5mm}
		F. Gursey, P. Ramond and P. Sikivie, \emph{A universal gauge theory model based on E6}, \href{https://doi.org/10.1016/0370-2693(76)90417-2}{Physics Letters B, 60(2), 177-180}; 
		\hspace{5mm}
		R. Barbieri, D. V. Nanopoulos and A. Masiero, \emph{Hierarchial fermion masses in E6}, \href{https://doi.org/10.1016/0370-2693(81)90589-X}{Physics Letters B, 104(3), 194-198}; 
		\hspace{5mm}
		G. Dvali and Q. Shafi, \emph{On proton stability and the gauge hierarchy problem}, \href{https://doi.org/10.1016/s0370-2693(97)00395}{Physics Letters B, 403(1-2), 65-69}.
		
		%%%%%
		
		\bibitem{32121DM} S. Bhattacharyya and A. Datta, \emph{Dark Matter perspective of Left-Right symmetric gauge model}, \href{https://doi.org/10.1016/j.nuclphysb.2023.116197}{Nucl. Phys. B, 991, 116197 (2023)}, arXiv: [\href{https://arxiv.org/pdf/2206.13105.pdf}{2206.13105}].
		
		%%%%%
		
		\bibitem{lrsm} N. G. Deshpande, J. F. Gunion, B. Kayser and F. Olness, \emph{Left-right-symmetric electroweak models with triplet Higgs field}, \href{https://doi.org/10.1103/PhysRevD.44.837}{Phys. Rev. D, 44, 837}; 
		\hspace{5mm}
		P. Duka, J. Gluza and M. Zralek, \emph{Quantization and renormalization of the manifest left-right symmetric model of electroweak interactions}, \href{https://doi.org/10.1006/aphy.1999.5988}{Annals Phys.280:336-408, 2000}, arXiv: [\href{https://arxiv.org/pdf/hep-ph/9910279.pdf}{hep-ph/9910279}].
		
		%%%%%
		
		\bibitem{heeck} J. Heeck, \emph{Phenomenology of Majorons}, \href{https://doi.org/10.3204/DESY-PROC-2017-02/heeck_julian}{DESY-PROC-2017-02 212-245, 2018}, arXiv: [\href{https://arxiv.org/pdf/1709.07670.pdf}{1709.07670}];
		\hspace{5mm}
		M. A. Diaz, M. A. Garcia-Jareno, D. A. Restrepo and J. W. F. Valle, \emph{Seesaw Majoron model of neutrino mass and novel signals in Higgs boson production at LEP}, \href{https://doi.org/10.1016/S0550-3213(98)00434-9}{Nucl. Phys. B 527, 44-60 (1998)}, arXiv: [\href{https://arxiv.org/pdf/hep-ph/9803362.pdf}{hep-ph/9803362}].
		
		%%%%%
		
		\bibitem{Atlas_h2} ATLAS Collaboration, \emph{Search for heavy neutral Higgs bosons produced in association with b-quarks and decaying to b-quarks at $\sqrt{s} = 13$ TeV with the ATLAS detector}, \href{https://doi.org/10.1103/PhysRevD.102.032004}{Phys. Rev. D 102, 032004 (2020)}, arXiv: [\href{https://arxiv.org/pdf/1907.02749.pdf}{1907.02749}].
		
		%%%%%
		
		\bibitem{Cms_h2} CMS Collaboration, \emph{Search for beyond the standard model Higgs bosons decaying into a $b\bar{b}$ pair in pp collisions at $\sqrt{s} = 13$ TeV}, \href{https://doi.org/10.1007/JHEP08(2018)113}{JHEP 08 (2018) 113}, arXiv: [\href{https://arxiv.org/pdf/1805.12191.pdf}{1805.12191}].
		
		%%%%%
		
		\bibitem{feynrules} A. Alloul, N. D. Christensen, C. Degrande, C. Duhr and B. Fuks, \emph{FeynRules 2.0 - A complete toolbox for tree-level phenomenology}, \href{https://doi.org/10.1016/j.cpc.2014.04.012}{Comput.Phys.Commun. 185 (2014) 2250-2300}, arXiv: [\href{https://arxiv.org/pdf/1310.1921.pdf}{1310.1921}].
		
		%%%%%
		
		\bibitem{madgraph} J. Alwall, R. Frederix, S. Frixione, V. Hirschi, F. Maltoni, O. Mattelaer \emph{et al.}, \emph{The automated computation of tree-level and next-to-leading order differential cross sections, and their matching to parton shower simulations}, \href{https://doi.org/10.1007/JHEP07(2014)079}{JHEP07 (2014) 079}, arXiv: [\href{https://arxiv.org/pdf/1405.0301.pdf}{1405.0301}].
		
		%%%%%
		
		\bibitem{parton-dist} R. D. Ball \emph{et al.}, \emph{Parton Distributions with LHC data}, \href{https://doi.org/10.1016/j.nuclphysb.2012.10.003}{Nucl. Phys. B867, 244 (2013)}, arXiv: [\href{https://arxiv.org/pdf/1207.1303.pdf}{1207.1303}].
		
		%%%%%
		
		\bibitem{h2-qcd-k-factor} S. Dawson, C. B. Jackson, L. Reina and D. Wackeroth, \emph{Higgs Production in Association With Bottom Quarks at Hadron Colliders}, \href{https://doi.org/10.1142/S0217732306019256}{Mod.Phys.Lett. A21 (2006) 89-110}, arXiv: [\href{https://arxiv.org/pdf/hep-ph/0508293.pdf}{hep-ph/0508293}];
		\hspace{5mm}
		O. Mattelaer, M. Mitra and R. Ruiz, \emph{Automated Neutrino Jet and Top Jet Predictions at Next-to-Leading-Order with Parton Shower Matching in Effective Left-Right Symmetric Models}, arXiv: [\href{https://arxiv.org/pdf/1610.08985.pdf}{1610.08985}].
		
		%%%%%
		
		\bibitem{b_running} A. V. Bednyakov, B. A. Kniehl, A. F. Pikelner and O. L. Veretin, \emph{On the b-quark running mass in QCD and the SM}, \href{https://doi.org/10.1016/j.nuclphysb.2017.01.004}{Nucl.Phys. B916 (2017) 463-483}, arXiv: [\href{https://arxiv.org/pdf/1612.00660.pdf}{1612.00660}].
		
		%%%%%
		
		\bibitem{AtlasCharged1} ATLAS Collaboration, \emph{Search for charged Higgs bosons decaying into a top quark and a bottom quark at $\sqrt{s} = 13$ TeV with the ATLAS detector}, \href{https://doi.org/10.1007/JHEP06(2021)145}{JHEP 06 (2021) 145}, arXiv: [\href{https://arxiv.org/pdf/2102.10076.pdf}{2102.10076}].
		
		%%%%%
		
		\bibitem{AtlasCharged2} ATLAS Collaboration, \emph{Search for charged Higgs bosons decaying into top and bottom quarks at $\sqrt{s} = 13$ TeV with the ATLAS detector}, \href{https://doi.org/10.1007/JHEP11(2018)085}{JHEP 11 (2018) 085}, arXiv: [\href{https://arxiv.org/pdf/1808.03599.pdf}{1808.03599}].
		
		%%%%%
		
		\bibitem{CmsCharged} CMS Collaboration, \emph{Search for charged Higgs bosons decaying into a top and a bottom quark in the all-jet final state of pp collisions at $\sqrt{s} = 13$ TeV}, \href{https://doi.org/10.1007/JHEP07(2020)126}{JHEP 07 (2020) 126}, arXiv: [\href{https://arxiv.org/pdf/2001.07763.pdf}{2001.07763}].
		
		%%%%%
		
		\bibitem{pythia8} T. Sjostrand, S. Mrenna and P. Z. Skands, \emph{PYTHIA 6.4 Physics and Manual}, \href{https://doi.org/10.1088/1126-6708/2006/05/026}{JHEP 0605:026,2006}, arXiv: [\href{https://arxiv.org/pdf/hep-ph/0603175.pdf}{hep-ph/0603175}].
		
		%%%%%
		
		\bibitem{delphes} DELPHES 3 Collaboration, J. de Favereau \emph{et al.}, \emph{A modular framework for fastsimulation of a generic collider experiment}, \href{https://doi.org/10.1007/JHEP02(2014)057}{J. High Energ. Phys. 2014, 57 (2014)},
		arXiv: [\href{https://arxiv.org/pdf/1307.6346.pdf}{1307.6346}].
		
		%%%%%
		
		\bibitem{tmva} A. Hoecker \emph{et al.}, \emph{TMVA - Toolkit for Multivariate Data Analysis}, arXiv: \href{https://doi.org/10.48550/arXiv.physics/0703039}{physics/0703039}.
		
	\end{thebibliography}
\end{document}